\documentclass[a4paper,11pt]{article}
\usepackage{amsfonts,amsmath,amsthm,amssymb}
\begin{document}
\title{Color-Flavor Transformation Revisited}
\author{Martin R. Zirnbauer\\
Institute for Theoretical Physics, University of Cologne,\\
Z\"ulpicher Stra{\ss}e 77a, 50937 K\"oln, Germany}
\date{September 21, 2021}

\maketitle

\begin{abstract}
The ``color-flavor transformation'', conceived as a kind of generalized Hubbard-Stratonovich transformation, is a variant of the Wegner-Efetov supermatrix method for disordered electron systems. Tailored to quantum systems with disorder distributed according to the Haar measure of a compact Lie group of any classical type ($A$, $B$, $C$, or $D$), it has been applied to Dyson's Circular Ensembles, random network models, disordered Floquet dynamical systems, quantum chaotic graphs, and more. We review the method and, in particular, explore its limits of validity. An application to $\mathrm{O}(N)$-Haar expectations of ratios of random characteristic polynomials is given. We also sketch a novel method to treat models where the color-flavor transformation fails.
\end{abstract}

\section{Introduction}

The study of disordered electron systems, a well-known prototype of which is the 3D Anderson tight-binding model with a random on-site potential, has led to rigorous results \cite{FS, AM} on wave function localization in the limit of strong disorder or small local density of states (LDoS). In the opposite limit of weak disorder or large LDoS, an elaborate theoretical-physics description of the metallic regime of extended states has been developed \cite{Efetov-Book, Wegner-Book}. What still remains open, to some extent, is the question of what is the precise nature of critical behavior at the Anderson localization-delocalization transition.

The tool box of theoretical techniques available for the study of disordered electrons includes numerical simulations and graphical expansions (a.k.a.\ the impurity diagram technique). While the latter do capture perturbative effects such as weak localization corrections and universal conductance fluctuations, much of the progress on higher-order perturbative and non-perturbative effects actually came from the adaptation of field-theoretical tools. The basic idea of the field-theory approach, pioneered by Wegner \cite{Wegner79, SW1980} and Efetov \cite{Efetov83}, is to trade the disorder average over i.i.d.\ random variables for an equivalent average over a certain supermatrix field. The latter offers the advantage of being correlated or collective (at least in a metallic regime), thereby inviting approximations of mean-field type followed by a systematic treatment of fluctuation corrections.

The Wegner-Efetov supermatrix method has been successful for model systems with a Hamiltonian, say
\begin{equation}
    H = H_d + H_f ,
\end{equation}
where $H_d$ is deterministic (e.g., a Laplacian for free-particle motion) and $H_f$ is fluctuating (e.g., an on-site random potential). The method starts out by expressing the retarded and advanced energy Green's functions $\langle \bullet \vert (E \pm \mathrm{i} \epsilon - H)^{-1} \vert \bullet^\prime \rangle$ as Gaussian field integrals over commuting and anti-commuting variables.
The disorder average is then taken inside the Gaussian field integral, where it yields the Fourier transform of the probability measure for the random variables in $H_f$. If the latter have a normal distribution, as is usually assumed, taking the Fourier transform results in a quartic interaction for the Gaussian fields. That interaction is brought back to a form quadratic in the Gaussian fields, by a so-called Hubbard-Stratonovich transformation introducing the Wegner-Efetov supermatrix field. By integrating out the Gaussian fields, one obtains an effective action for the latter. Saddle-point analysis followed by a gradient expansion (valid in a regime of metallic or locally diffusive behavior) leads to a field theory of the type of a nonlinear $\sigma$ model.

In the present article, we are concerned with a somewhat different class of model systems, where instead of the continuous-time dynamics generated by a Hamiltonian $H$ one has a discrete-time dynamics generated by a unitary operator, $U$. Concretely, we will consider models where $U$ is a product of unitary operators
\begin{equation}
    U = U_d U_f ,
\end{equation}
with $U_d$ deterministic and $U_f$ fluctuating. To make our models amenable to analytical treatment, we take the random variables in $U_f$ to be distributed according to the Haar measure of a compact Lie group (e.g.\ $\mathrm{U}(1)^{N}$ for a system with $N$ lattice sites). Examples of such models are Dyson's Circular Ensembles (where $U_d \equiv 1$ is trivial), the random-phase quantum kicked rotor, models akin to the Chalker-Coddington network model, etc.

Now in the early going, it was not clear how to adapt the Wegner-Efetov approach to that kind of model. Let us give some indication why. If $\vert j \rangle$ denotes an orthonormal basis for the Hilbert space of our quantum system --- typically a tight-binding or Wannier basis in the case of disordered electrons --- the retarded Green's function for the unitary operator $U$ is a collection of matrix elements
\begin{equation}
    \langle j \vert (1- \zeta U)^{-1} \vert k \rangle \,,
\end{equation}
where a resolvent parameter $\zeta \in \mathbb{C}$ with $|\zeta| < 1$ is inserted to make the geometric series $(1- \zeta U)^{-1} = \sum_{n \geq 0} \zeta^n U^n$ converge. The corresponding matrix element of the advanced Green's function is
\begin{equation}
    \langle k \vert (1- \bar\zeta U^{-1})^{-1} \vert j \rangle = \overline{\langle j \vert (1- \zeta U)^{-1} \vert k \rangle} \,,
\end{equation}
with convergent series expansion $(1- \bar\zeta U^{-1})^{-1} = \sum_{n \geq 0} \bar\zeta^n U^{-n}$. The standard trick now is to introduce two sets (indexed by $\nu = 0, 1$) of complex variables $\varphi^\nu (i)$, commuting for $\nu = 0$ and anti-commuting for $\nu = 1$, in order to express the retarded Green's function as a Gaussian integral:
\begin{equation}
    \langle j \vert (1- \zeta U)^{-1} \vert k \rangle = \int_{\varphi,\bar\varphi} \mathrm{e}^{-
    \bar{\varphi}_\nu \cdot (1- \zeta U) \varphi^\nu} \varphi^0 (j) \bar\varphi_0 (k) ,
\end{equation}
where
\begin{equation}
    \bar{\varphi}_\nu \cdot (1- \zeta U) \varphi^\nu \equiv \sum\nolimits_{i,i^\prime} \bar{\varphi}_\nu(i)\, \langle i\vert (1-\zeta U) \vert i^\prime \rangle\, \varphi^\nu (i^\prime),
\end{equation}
and $\int_{\varphi,\bar\varphi}$ means that we integrate with Lebesgue measure (suitably normalized to absorb a constant) for the commuting variables times the product of partial derivatives for all the anti-commuting variables. We are assuming the summation convention for the even-odd index $\nu$. The advanced Green's function is expressed in the same way, by introducing another set of complex variables $\psi^\nu (i)$:
\begin{equation}
    \langle k \vert (1- \bar\zeta U^{-1})^{-1} \vert j \rangle  = \int_{\psi,\bar\psi} \mathrm{e}^{- \psi^\nu \cdot (1-\bar\zeta U^{-1}) \bar\psi_\nu (-1)^{\nu}} \bar\psi_0 (k) \psi^0 (j) .
\end{equation}
(Please be advised that the fermion parity factor $(-1)^\nu$ and the varying position of the index $\nu$ will be instrumental in building a unified framework to handle all disorder types.)

Let us now specialize to the simple example of $U = U_d U_f$ with diagonal $U_f$,
\begin{equation}
    \langle i \vert U_f \vert i^\prime \rangle = \delta_{i^\prime}^{i} \, \mathrm{e}^{ \mathrm{i} \theta_i} ,
\end{equation}
parameterized by random phases $\theta_i$ with probability measure $d\theta_i / 2\pi$, a.k.a.\ Haar measure on $\mathrm{U}(1)$. In that setting, when we take the disorder average of a product of retarded and advanced Green's functions, the random-phase integral results in a product of Bessel functions:
\begin{equation}
    \prod_i \int \frac{d\theta_i}{2\pi} \exp \left( \mathrm{e}^{\mathrm{i} \theta_i} \zeta X(i) + \mathrm{e}^{- \mathrm{i}\theta_i} \bar\zeta Y(i) \right) = \prod_i I_0 \left( |\zeta|^2 X(i) Y(i) \right) ,
\end{equation}
where
\begin{equation}
    X(i) = \sum_j \bar\varphi_\nu(j)\, \langle j \vert U_d \vert i \rangle\, \varphi^\nu(i) , \quad Y(i) = \sum_j \bar\psi_\nu(j)\, \langle i \vert U_d^{-1} \vert j \rangle\, \psi^\nu(i) .
\end{equation}
Clearly, the integrand's Gaussian dependence on $\varphi$ and $\psi$ has become very nonlinear, making further progress difficult. Here is where the so-called ``color-flavor transformation'' comes in. Based on the principle of Howe duality \cite{Howe-Trans, Howe-Schur} for Fock representations of particles carrying color and flavor, it restores the Gaussian dependence on the Gaussian fields $\varphi$ and $\psi$, by introducing a supermatrix field analogous to the Hubbard-Stratonovich field of the standard Wegner-Efetov approach. Further analysis then proceeds as usual: one integrates out the Gaussian field and develops an effective theory for the supermatrix field. The reader is referred to the body of the paper for detailed statements of the color-flavor transformation; see Eqs.\ (\ref{eq:type-D}), (\ref{eq:type-C}), and (\ref{eq:type-A}) below.

In this introductory section, we shall visit a couple of illustrative special cases, as these will help us indicate some mathematical background and appreciate the limits of validity of the color-flavor transformation. Let then $g$ be a random orthogonal $N \times N$ matrix with probability distribution given by the unit-mass Haar measure $dg$ on $\mathrm{O}(N)$ --- a case not covered by the original reference \cite{circular}. We might be interested in the Haar expectation of a product of $n$ characteristic polynomials, $\prod_{\mu=1}^n \mathrm{Det}(\alpha_\mu - g)$, depending on complex parameters $\alpha_\mu\,$. This can be processed by expressing each determinant as a Gaussian integral over anti-commuting variables, then applying the color-flavor transformation (or, rather, a ``fermionic'' version thereof), and finally integrating out the anti-commuting variables to produce another product of determinants (actually, Pfaffians). The result is
\begin{eqnarray}\label{eq:typeBD-FF}
    &&\int_{\mathrm{O}(N)} \!\!\!\! dg \, \prod_{\mu=1}^n \mathrm{Det}(\alpha_\mu - g) \cr &&= c_{N,\,n}^{(1)} \int_{M_1} D(Z,\bar{Z}) \, \mathrm{Det}^{-n+1-N/2} \begin{pmatrix} 1 &Z \cr \bar{Z} &1 \end{pmatrix} \mathrm{Det}^{N/2} \begin{pmatrix} \alpha &Z\cr \bar{Z} &\alpha\end{pmatrix},
\end{eqnarray}
where $\alpha = {\rm diag}(\alpha_1, \ldots, \alpha_n)$ is a diagonal matrix, and the integral on the right-hand side is over the domain, $M_1$, of all complex skew $n \times n$ matrices $Z = - Z^T$, with integration measure $D(Z,\bar{Z})$ given by the product of Lebesgue measures for all independent matrix elements (i.e., the $Z^{\mu\nu}$ for $\mu < \nu$) of $Z$. The symbol $\bar{Z}$ means the complex conjugate of $Z$.

We remark that with the constant $c_{N,\,n}^{(1)}$ chosen correctly, Eq.\ (\ref{eq:typeBD-FF}) holds for \underline{any} pair of positive integers $N, n$. As a matter of fact, Eq.\ (\ref{eq:typeBD-FF}) can be seen as a direct consequence of the Borel-Weil theory \cite{BorelWeil, stable} of irreducible highest-weight representations of a compact Lie group -- in the present instance: $\mathrm{SO} (2n)$-representations with highest weight of $\mathrm{U}(n)$-scalar type indexed by $N$. [To be precise, these representations are double-valued when $N$ is odd; they are single-valued as representations of a $2:1$ cover $\mathrm{Spin}(2n) \to \mathrm{SO}(2n)$.] Mathematically speaking, that theory tells us that the vectors of the said representations can be represented as holomorphic sections of a complex line bundle over the compact Hermitian symmetric space $M_1 \cong \mathrm{SO}(2n) / \mathrm{U}(n)$. In the language of physics, such vectors have a representation by generalized spin-coherent states parameterized by $Z$.

We turn to our second illustration: the $\mathrm{O}(N)$-Haar expectation of a product of reciprocal characteristic polynomials, $\prod_\mu \mathrm{Det}^{-1}(\alpha_\mu - g)$. To avoid pole singularities, we must now restrict the range of the complex parameters $\alpha_\mu\,$, say by $|\alpha_\mu| > 1$. Proceeding in the same way as before, but using commuting instead of anti-commuting integration variables, we deduce by the (``bosonic'' version of the) color-flavor transformation the following identity:
\begin{eqnarray}\label{eq:typeBD-BB}
    &&\int_{\mathrm{O}(N)} \!\!\!\! dg \, \prod_{\mu=1}^n \mathrm{Det}^{-1}(\alpha_\mu - g) \cr &&= c_{N,\,n}^{(0)} \int_{M_0} D(Z,\bar{Z}) \, \mathrm{Det}^{-n-1+N/2} \begin{pmatrix} 1 &Z \cr \bar{Z} &1 \end{pmatrix} \mathrm{Det}^{-N/2} \begin{pmatrix} \alpha &Z\cr \bar{Z} &\alpha\end{pmatrix} .
\end{eqnarray}
Here the $n \times n$ matrix $Z$ is complex symmetric, and the domain of integration $M_0$ is the noncompact domain $\bar{Z} Z < 1$ (a.k.a.\ a classical bounded symmetric domain). Note also the sign changes $N \to -N$ and $n-1 \to n + 1$. An important piece of information now is this: the formula (\ref{eq:typeBD-BB}), as it stands, holds true only in the range of $N \geq 2n+1$. That restriction is, in fact, not difficult to verify: the power $-n-1+N/2$ of the middle determinant must not be below $-1/2$ if the integral over $M_0$ is to converge.

The mathematical object in the background here is a discrete series of unitary representations of the noncompact group $\mathrm{Sp}(2n,\mathbb{R})$; more precisely, of infinite-dimensional irreducible representations of the metaplectic group $\mathrm{Mp}(2n)$ [a $2:1$ cover of $\mathrm{Sp}(2n,\mathbb{R})$], with the highest-weight vector carrying a one-dimensional representation of $\mathrm{U}(n)$. (The latter is double-valued if $N$ is odd. The $N=1$ representation of $\mathrm{Mp}(2n)$ is known as the Segal-Shale-Weil representation in mathematics, and as the oscillator representation, or bosonic Fock space, in physics.) Such representations come with an $\mathrm{Sp}$-invariant Hermitian scalar product \cite{RossiVergne, KashiVergne}, and their state vectors are still in one-to-one correspondence with holomorphic sections of a complex line bundle, now over a symmetric space of noncompact type, namely $M_0 \cong \mathrm{Sp}(2n,\mathbb{R}) / \mathrm{U}(n)$. Our formula (\ref{eq:typeBD-BB}) can be deduced if the $\mathrm{Sp}$-invariant Hermitian scalar product has an $L^2$-realization by integration over $M_0\,$. The latter criterion amounts \cite{stable} to the stated inequality, $N \geq 2n+1$.

In the present article, we will focus on the ``super''-version of Eqs.\ (\ref{eq:typeBD-FF}, \ref{eq:typeBD-BB}) and related identities -- mixing commuting with anti-commuting variables, or bosons with fermions -- as that version is the one needed to tackle the disordered-electron challenge of computing expectations of products of retarded and advanced single-electron Green's functions. We will review the color-flavor transformation for the full list of cases with Haar-distributed disorder on one of the classical compact Lie groups; namely, the unitary group $\mathrm{U}(N)$, the compact symplectic group $\mathrm{Sp}(N)$, and the real orthogonal group $\mathrm{O}(N)$; these are known as type $A$, type $C$, and type $BD$, respectively.

Our main concern in the present paper is with validity questions, which were left open by the existing literature. We will show that the color-flavor transformation in its standard form is valid as long as the number of ``colors'' ($N$) lies in a stable range above a threshold set by the number of bosonic ``flavors'' ($n$). Here it should be observed that a closely related condition on range is known to guarantee the validity of the superbosonization formula \cite{LSZ}. It should also be mentioned that Fyodorov and Khoruzhenko \cite{FK} have extended the bosonic version of the type-$A$ color-flavor transformation from its stable range $N \geq 2n$ to the range of $2n > N \geq n$; a similar extension is expected to be possible for the types $C$ and $BD$. Unfortunately, it remains unknown, in general, how to go beyond the stable range in the super-case. For that, one would need a well-developed analog of Borel-Weil theory (for the relevant supergroup representations) and analytical mastership of the matrix coefficients and, especially, the question of their $L^2$-type integrability. Since such a theory does not seem to be available (see, however, \cite{Shibata}), we have to make do with case-by-case considerations. In particular, we will study the case of $\mathrm{U}(N)$ color and a single flavor of retarded and advanced bosons and fermions ($n=1$). We will demonstrate that, there, the standard form of the color-flavor transformation fails for $N=1$. We will also show how to correct the formula in that special case. Since the formula of the corrected transformation turns out to be rather unwieldy, we go on to develop an alternative formula with better prospects for practical applicability.

A summary of the contents of the paper is as follows. In Section \ref{sect:WhatIs} we state the color-flavor transformation for all types in the sequence of $BD$, $C$, $A$, and we give a quick application of the type-$BD$ transformation to $\mathrm{O}(N)$-Haar expectations of ratios of random characteristic polynomials. Section \ref{sect:origin} explains the mathematics behind the color-flavor transformation, starting from its origin in invariant theory, viz.\ the principle of the Howe duality for Fock representations. Our proof strategy is sketched in some detail for the case of type $BD$. Section \ref{sect:TPfctn} is an excursion to draw attention to the difficulties that arise when one wants to apply the color-flavor transformation outside the stable range of a large number of colors; this is highlighted at the basic example of two-point functions for the case of $\mathrm{U}(N)$-Haar disorder, focusing especially on $N = 1$. In Section \ref{sect:bounds} we give sufficient conditions for the color-flavor transformation to hold in standard form.

\section{What is the color-flavor transformation?}\label{sect:WhatIs}

Understood in the sense of the present paper, the color-flavor transformation \cite{circular} is a variant of the Wegner-Efetov supermatrix method \cite{Wegner79, SW1980, Efetov83}, a tool widely used in the field of random matrices and disordered noninteracting electrons. The standard version of the supermatrix method applies to systems with continuous-time quantum dynamics generated by a Hermitian operator (the Hamiltonian). Its core step is a so-called Hubbard-Stratonovich transformation, trading the average over the disorder for an average over a supermatrix field. In comparison, the color-flavor transformation, which is of a similar nature and scope, albeit less well known, applies to systems with discrete-time quantum dynamics generated by a unitary operator (e.g., the Floquet operator of a quantum Hamiltonian system with periodic driving). While the standard supermatrix method is most natural for systems with disorder given by Gaussian random variables, the color-flavor transformation is most natural for systems with disorder given by the Haar measure on a compact classical Lie group. Correspondingly, the latter exists in the form of three different types that are listed below.
\begin{itemize}
\item The color-flavor transformation of type $A$ applies to Dyson's Circular Unitary Ensemble $\mathrm{CUE}_N$ \cite{Dyson-CE} and models derived from it (e.g., by taking tensor products).
\item Type $C$ applies to models with disorder distributed according to Haar measure on the group $\mathrm{Sp}(N) \equiv \mathrm{USp}(N)$ (with even $N$) of unitary symplectic transformations. Such models appear, e.g., in the description of superconducting or superfluid systems (where the $\mathrm{U}(1)$ phase-rotation symmetry is spontaneously broken due to Cooper pair formation) with conserved spin and broken time-reversal symmetry.
\item Type $BD$ is analogous to type $C$ except that the symplectic group $\mathrm{Sp}(N)$ is replaced the orthogonal group $\mathrm{O}(N)$. Type $BD$ splits into two subtypes, denoted by $B$ for $N$ odd and $D$ for $N$ even. Applications (of subtype $D$, by the Bogoliubov-deGennes formalism of Hartree-Fock-Bogoliubov mean-field theory) exist for superconductors or superfluids without spin rotation and time-reversal symmetry.
\end{itemize}
From the notational perspective, the color-flavor transformation of type $BD$ turns out to be the simplest one. For that reason, and also motivated by the circumstance that $BD$ was not included in the original reference \cite{circular}, we begin with that type.

\subsection{Color-flavor transformation of type $BD$}\label{sect:2.1}

Here we motivate and state the color-flavor (CF) transformation of type $BD$, leaving the mathematical background and proof for a later section. The setup is as follows.

i.\ We introduce variables $\psi_{\;i}^\mu$ with indices $i = 1, \ldots, N$ for ``color'' and $\mu = 1, \ldots, n$ for ``flavor''. Depending on the flavor index, the $\psi_{\;i}^\mu$ are commuting or anti-commuting (a.k.a.\ Grassmann) variables. In the former case, the so-called fermion parity is set to $|\mu| = 0$, in the latter case $|\mu| = 1$. We also write $n = n_0 + n_1$, where $n_0$ and $n_1$ are the numbers of flavors with even and odd fermion parity, respectively. We remark that in the Wegner-Efetov supermatrix method for disordered electron systems, one takes the same number of commuting and anti-commuting variables ($n_0 = n_1$), but some other choices also have interesting applications; for example, to compute determinant-determinant correlation functions, one takes $n_1 \not= n_0 = 0$.

ii.\ We augment the set of variables $\psi_{\;i}^\mu$ with conjugate variables $\bar\psi_{\;\mu}^i$ of the same index range and fermion parity. Pairing $\psi$'s with $\bar\psi$'s, we form quadratic arrays $\bar\psi_{\;\mu}^i \psi_{\;j}^\mu$ in color space, which are then coupled to orthogonal $N \times N$ matrices $g \in \mathrm{O}(N)$:
\begin{equation}\label{eq:ginv}
    (g^{-1})_{\;i}^j \bar\psi_{\;\mu}^i \psi_{\;j}^\mu \,.
\end{equation}
Here and throughout this paper, the summation convention is in force.

iii.\ Let $dg$ denote the Haar measure for $\mathrm{O}(N)$ with total mass one ($\int_{\mathrm{O}(N)} dg = 1$) and consider the integral
\begin{equation}\label{eq:CF-LHS}
    \Omega \equiv \int_{\mathrm{O}(N)} \!\!\!\! dg \; \exp \left( (g^{-1})_{\;i}^j \bar\psi_{\;\mu}^i \psi_{\;j}^\mu \right) .
\end{equation}
This is the expression on the left-hand side of the CF transformation (of type $BD$). (For the statement of Eq.\ (\ref{eq:type-D}) below, we replace $g^{-1} \to g$ by invariance of the Haar measure, but for reasons that will become clear later, we prefer to leave it here as it stands.)

iv.\ Next, we make a pedagogical effort to motivate the right-hand side of the CF transformation, as follows. By definition, the action of the orthogonal group $\mathrm{O}(N)$ on the Euclidean vector space $\mathbb{R}^N$ preserves the symmetric bilinear form which is given by the components, $\delta_{ij} = \delta_{ji}\,$, of the Euclidean metric on $\mathbb{R}^N$ (in any basis). If $\delta^{ij}$ are the components of the dualized (or inverse) bilinear form, one has
\begin{equation}
    g_{\;k}^i \delta_{ij} g_{\;l}^j = \delta_{kl} \,, \quad
    g_{\;i}^k \delta^{ij} g_{\;j}^l = \delta^{kl} ,
\end{equation}
for $g \in \mathrm{O}(N)$. We now expand the exponential in the integral (\ref{eq:CF-LHS}) and compute the first few terms of the resulting series (with $g^{-1}$ replaced by $g$). Performing some basic integrals,
\begin{equation}
    \int_{\mathrm{O}(N)} \!\!\!\! dg \; g_{\; j}^i = 0 , \quad \int_{\mathrm{O}(N)} \!\!\!\! dg \; g_{\; k}^i g_{\; l}^j = N^{-1} \delta^{ij} \delta_{lk} \,,
\end{equation}
we then obtain
\begin{equation}\label{eq:Omega1}
    \Omega = 1 + (2N)^{-1} (\bar\psi_{\;\mu}^i \psi_{\;j}^\mu) \delta^{jk} (\bar\psi_{\;\nu}^l \psi_{\;k}^\nu) \delta_{li} + \mathcal{O}(N^{-2}) .
\end{equation}
Next, we reorder the second pair: $\bar\psi_{\;\nu}^l \psi_{\;k}^\nu = \psi_{\;k}^\nu (-1)^{|\nu|} \bar\psi_{\;\nu}^l\,$, where the sign factor $(-1)^{|\nu|}$ appears because both $\bar\psi_{\; \nu}^l$ and $\psi_{\;k}^\nu$ are Grassmann variables if $|\nu| = 1$. As a follow-up, we make a cyclic rearrangement to change the coupling scheme from flavor-invariant to color-invariant pairs:
\begin{equation}\label{eq:rearr}
    (\bar\psi_{\;\mu}^i \psi_{\;j}^\mu) \delta^{jk} (\bar\psi_{\;\nu}^l \psi_{\;k}^\nu) \delta_{li}     = (-1)^{|\mu|} (\psi_{\;j}^\mu \delta^{jk} (-1)^{|\nu|} \psi_{\;k}^\nu) (\bar\psi_{\;\nu}^l \delta_{li} \bar\psi_{\;\mu}^i) .
\end{equation}
The logic behind the positioning of the sign factors is this: in the outer sum over $\mu$, the presence of $(-1)^{|\mu|}$ accounts for the operation of ``supertrace'', which is the natural trace to take here; in the inner sum over $\nu$, the presence of $(-1)^{|\nu|}$ reflects an operation of ``supertranspose'' (indeed, the indices on $\psi_{\;k}^\nu$ and $\bar\psi_{\;\nu}^l$ are out of order for consistent matrix multiplication, and reversing the index order by supertransposition brings about a sign factor for $\psi_{\;k}^\nu$). Note the exchange symmetries
\begin{equation}
    Q^{\mu\nu} \equiv \psi_{\;j}^\mu \delta^{jk} (-1)^{|\nu|} \psi_{\;k}^\nu = (-1)^{|\mu| |\nu| + |\mu| + |\nu|} Q^{\nu\mu}
\end{equation}
and
\begin{equation}
    \widetilde{Q}_{\nu\mu} \equiv \bar\psi_{\;\nu}^l \delta_{li} \bar\psi_{\;\mu}^i = (-1)^{|\mu| |\nu|} \widetilde{Q}_{\mu\nu} \,.
\end{equation}
With the abbreviations $Q$ and $\widetilde{Q}$, the expression (\ref{eq:Omega1}) for $\Omega$ takes the form
\begin{equation}\label{eq:Omprime}
    \Omega = 1 + (2N)^{-1} (-1)^{|\mu|} Q^{\mu\nu} \widetilde{Q}_{\nu\mu} +  \mathcal{O}(N^{-2}) .
\end{equation}

v.\ In the next step, we introduce in flavor space two square complex supermatrices $Z, \widetilde{Z}$ modeled after $Q, \widetilde{Q}$:
\begin{equation}\label{eq:exch-Z}
    Z^{\mu\nu} = (-1)^{|\mu| |\nu| + |\mu| + |\nu|} Z^{\nu\mu} , \quad \widetilde{Z}_{\nu\mu} = (-1)^{|\mu| |\nu|} \widetilde{Z}_{\mu\nu} \,.
\end{equation}
By decree (within the confines of this pedagogical introduction), the matrix elements of $Z, \widetilde{Z}$ are to behave as Gaussian-distributed variables with vanishing first moments,
\begin{equation}\label{eq:1st-mom}
    \langle Z^{\mu\nu} \rangle = \langle \widetilde{Z}_{\nu\mu} \rangle = 0 ,
\end{equation}
and second moments
\begin{eqnarray}\label{eq:2nd-mom}
    &&\langle Z^{\mu\nu} Z^{\lambda\rho} \rangle = \langle \widetilde{Z}_{\mu\nu} \widetilde{Z}_{\lambda\rho} \rangle = 0, \cr &&\langle Z^{\mu\nu} \widetilde{Z}_{\lambda\rho} \rangle = \frac{1}{N} (-1)^{|\nu|} \left( \delta^{\nu}_{\lambda} \delta^{\mu}_{\rho} + (-1)^{|\mu| |\nu|} \delta^{\mu}_{\lambda} \delta^{\nu}_{\rho} \right) .
\end{eqnarray}
We couple these matrices to the color invariants on the right-hand side of Eq.\ (\ref{eq:rearr}) and take the expectation of the exponentiated sum:
\begin{equation}\label{eq:omegap}
    \Omega^\prime \equiv \left\langle \exp \left( {\textstyle{\frac{1}{2}}} (-1)^{|\mu|} Z^{\mu\nu}
    \widetilde{Q}_{\nu\mu} + {\textstyle{\frac{1}{2}}} (-1)^{|\nu|} \widetilde{Z}_{\nu\mu} Q^{\mu\nu} \right) \right\rangle .
\end{equation}
We again expand the exponential and compute the first few terms of the resulting series in $1/N$, by using the expressions (\ref{eq:1st-mom}, \ref{eq:2nd-mom}) for the moments of $Z$ and $\widetilde{Z}$. It is readily seen that the leading terms agree with those of $\Omega$ in (\ref{eq:Omprime}):
\begin{equation}
    \Omega^\prime = 1 + (2N)^{-1} (-1)^{|\nu|} \widetilde{Q}_{\nu\mu} Q^{\mu\nu} + \mathcal{O}(N^{-2}) .
\end{equation}
The observation $\Omega = \Omega^\prime$ (up to terms of order $N^{-2}$) motivates the following development.

vi.\ To turn the approximate equality $\Omega \approx \Omega^\prime$ into an exact identity, one needs to make a ``quantum deformation'' of the Gaussian expectation (\ref{eq:omegap}). To that end, we decompose the supermatrices $Z$ and $\widetilde{Z}$ by blocks,
\begin{equation}
    Z = \begin{pmatrix} Z^{\rm BB} &Z^{\rm BF} \cr Z^{\rm FB} &Z^{\rm FF} \end{pmatrix} , \quad \widetilde{Z} = \begin{pmatrix} \widetilde{Z}_{\rm BB} &\widetilde{Z}_{\rm BF} \cr \widetilde{Z}_{\rm FB} &\widetilde{Z}_{\rm FF} \end{pmatrix} ,
\end{equation}
and take the integration domain, $M_0\,$, for the even-even (or boson-boson) blocks $Z^{\rm BB}, \widetilde{Z}_{\rm BB}$ to be
\begin{equation}\label{eq:restr}
    \widetilde{Z}_{\rm BB} = + (Z^{\rm BB})^\dagger , \quad 0 \leq Z^{\rm BB} \widetilde{Z}_{\rm BB}  < 1 ,
\end{equation}
where $\dagger$ denotes the Hermitian adjoint. The integration domain, $M_1\,$, for the odd-odd (or fermion-fermion) blocks $Z^{\rm FF}, \widetilde{Z}_{\rm FF}$ is taken to be
\begin{equation}\label{eq:Herm-FF}
    \widetilde{Z}_{\rm FF} = - (Z^{\rm FF})^\dagger , \quad 0 \leq - Z^{\rm FF} \widetilde{Z}_{\rm FF}  < \infty .
\end{equation}
Further, let $D\mu(Z,\widetilde{Z})$ be a Berezin integral form that results by standard construction \cite{Berezin} from the metric tensor
\begin{equation}\label{eq:inv-metric}
    \mathrm{STr}\, (1 - Z \widetilde{Z})^{-1} d{Z} \, (1 - \widetilde{Z} Z)^{-1} d\widetilde{Z} \,
\end{equation}
with $\mathrm{STr}\, A = (-1)^{|\mu|} A_{\ \mu}^\mu$ the supertrace. The result of that construction is
\begin{equation}\label{eq:invt-Dm}
    D\mu(Z,\widetilde{Z}) = D(Z,\widetilde{Z}) \circ \mathrm{SDet}^{-n_0 + n_1 - 1}(1 - \widetilde{Z} Z) ,
\end{equation}
where the symbol $\circ$ means composition of operators, $\mathrm{SDet}$ is the superdeterminant, and $D(Z,\tilde{Z})$ denotes the so-called ``flat measure'', i.e.\ the product of all independent differentials (of the even variables) and independent partial derivatives (w.r.t.\ the odd variables):
\begin{equation}\label{eq:flat-D}
    D(Z,\widetilde{Z}) = \prod_{|\mu| + |\nu| \; {\rm even}} d Z^{\mu\nu} \, d\widetilde{Z}_{\nu\mu} \times \prod_{|\mu| + |\nu| \; {\rm odd}} \frac{\partial^2}{\partial Z^{\mu\nu} \, \partial \widetilde{Z}_{\nu\mu}} \,.
\end{equation}
Furthermore, let $c_N^{(BD)}$ be a normalization constant determined by
\begin{equation}\label{eq:cN}
    c_N^{(BD)}\int_{M_0 \times M_1} \!\!\! D\mu (Z,\widetilde{Z}) \, \mathrm{SDet}^{N/2} (1-\widetilde{Z} Z) = 1 .
\end{equation}
[Please be warned that the constant $c_N^{(BD)}$ depends not only on $N$ but also on $n_0$ and $n_1$. We omit the latter from our notation by the rationale that the superscript $(BD)$ should suffice as a reminder of that dependence.] We then claim that the following identity,
\begin{eqnarray}\label{eq:type-D}
    &&\int_{\mathrm{O}(N)} \!\!\! dg \, \exp \left(  g_{\; j}^i \bar\psi_{\;\mu}^j \psi_{\; i}^\mu \right) = c_N^{(BD)} \int_{M_0 \times M_1} \!\!\!\!\!\!\! D\mu (Z,\widetilde{Z}) \; \mathrm{SDet}^{N/2}(1 - \widetilde{Z} Z) \\ &&\hspace{4cm} \times \exp \left( {\textstyle{\frac{1}{2}}} \delta^{jk} \psi_{\;k}^\nu (-1)^{|\nu|} \widetilde{Z}_{\nu\mu} \psi_{\;j}^\mu + {\textstyle{\frac{1}{2}}} \bar\psi_{\;\mu}^i Z^{\mu\nu} \bar\psi_{\;\nu}^l \delta_{li} \right) , \nonumber
\end{eqnarray}
referred to as the CF transformation of type $BD$, holds with a finite constant $c_N^{(BD)}$ whenever $N \geq 2 n_0 + 1$. (Recall that $n_0$ is the number of bosonic flavors.)

\medskip\noindent\textbf{Remark 1.} From the perspective of applications, the merit of the identity (\ref{eq:type-D}) is that it transforms from an expression that is color-coupled but diagonal (hence uncoupled) in the flavor index, to an expression where the situation is reversed: the exponent on the right-hand side is flavor-coupled but color-uncoupled.

\medskip\noindent\textbf{Remark 2.} It should be stressed that the integral on the left-hand side of the CF transformation (\ref{eq:type-D}) is over $\mathrm{O}(N)$ (with two connected components from $\mathrm{Det} \, g = +1$ and $\mathrm{Det}\, g = - 1$), \underline{not} the special orthogonal group $\mathrm{SO}(N)$. The analogous trans\-formation for $\mathrm{SO}(N)$ is more complicated. The reason is that all invariant tensors for $\mathrm{O}(N)$ are polynomials in the basic invariant $\delta_{ij}\,$, whereas $\mathrm{SO}(N)$ has an additional invariant $\varepsilon_{i_1 i_2 \cdots \, i_N}$ (the totally anti-symmetric epsilon tensor for $\mathbb{R}^N$). The latter gives rise to corrections to the right-hand side of Eq.\ (\ref{eq:type-D}) when $\mathrm{O}(N)$ is replaced by $\mathrm{SO}(N)$.

\medskip\noindent\textbf{Remark 3.} From the mathematical perspective of differential geometry, the matrix elements of $Z$ and $\widetilde{Z}$ are local coordinates for a Hermitian symmetric superspace $G/K$ of type $C{\rm I}|D{\rm I\!I\!I}$ \cite{suprev} with K{\"a}hler superpotential $\mathrm{STr} \ln (1-\widetilde{Z} Z)$. The underlying symmetric spaces are $M_0 = \mathrm{Sp}(2n_0,\mathbb{R}) / \mathrm{U}(n_0)$ and $M_1 = \mathrm{SO}(2n_1) / \mathrm{U}(n_1)$, where we recall that $n_0$ (resp.\ $n_1$) is the number of even (resp.\ odd) flavors.

\medskip\noindent\textbf{Remark 4.} Eq.\ (\ref{eq:type-D}) includes two extreme cases: $n_0 = 0$ and $n_1 = 0$; in the former case we speak of the ``fermionic'' version of the CF transformation, in the latter case of the ``bosonic'' version. Both of these are easy corollaries of results in classical mathematics.

\subsection{An application}\label{sect:WeylCF}

We now put the theoretical development briefly on hold and illustrate the CF trans\-formation of type $BD$ by working through a simple example: Haar expectations of ratios of characteristic polynomials for the orthogonal group -- these received much attention in the late 1990's, around the time when the CF transformation was conceived.

For a set of complex parameters $\alpha_1 , \ldots, \alpha_n$ we consider
\begin{equation}
    \Omega(\alpha) = \int\limits_{\mathrm{O}(N)} \!\! dg \ \frac{\prod_{\nu > n_0} \mathrm{Det}(\alpha_\nu - g)}{\prod_{\mu \leq n_0} \mathrm{Det} (\alpha_\mu - g)} \,,
\end{equation}
which may serve as a generating function for numerous observables including Det-Det correlations and multi-level correlation functions. To carry out the integral over the group $\mathrm{O}(N)$ with Haar measure $dg$, we express the determinants and their reciprocals as Gaussian integrals over anti-commuting and commuting variables, respectively:
\begin{equation}
    \frac{\prod_{\nu > n_0} \mathrm{Det}(\alpha_\nu - g)}{\prod_{\mu \leq n_0} \mathrm{Det} (\alpha_\mu - g)} = \int_{\psi,\,\bar\psi} \exp \left( g_{\; j}^i \bar\psi_{\;\lambda}^j \psi_{\; i}^\lambda - \bar\psi_{\;\lambda}^i \alpha_{\; \rho}^\lambda \psi_{\; i}^\rho \right) ,
\end{equation}
where $\alpha_{\; \rho}^{\lambda}$ are the matrix elements of the diagonal supermatrix $\alpha$ with diagonal entries $\alpha_\lambda\,$. (For convergence of the integral, we must assume $\mathrm{Re}\, \alpha_\mu > 1$ for $\mu = 1, \ldots, n_0\,$. If that restriction is undesirable for the particular observable to be computed, it can easily be circumvented by variable substitution or analytic continuation.) We are now ready to apply the CF transformation (\ref{eq:type-D}). Changing the order of integration, we obtain
\begin{eqnarray}
    &&\Omega(\alpha) = c_N^{(BD)} \int_{M_0 \times M_1} \!\!\!\!\!\! D\mu (Z,\widetilde{Z}) \; \mathrm{SDet}^{N/2}(1 - \widetilde{Z} Z) \\ &&\hspace{1cm} \times \int_{\psi,\, \bar\psi} \exp \left( {\textstyle{\frac{1}{2}}} \delta^{ij} \psi_{\;j}^\rho (-1)^{|\rho|} \widetilde{Z}_{\rho\lambda} \psi_{\;i}^\lambda + {\textstyle{\frac{1}{2}}} \bar\psi_{\;\lambda}^i Z^{\lambda\rho} \bar\psi_{\;\rho}^j \delta_{ji} - \bar\psi_{\;\lambda}^i \alpha_{\; \rho}^{\lambda} \psi_{\; i}^\rho \right) . \nonumber
\end{eqnarray}
The beautiful feature due to (\ref{eq:type-D}) is that the integral over the variables $\psi, \bar\psi$ factors in the color index $i = 1, ..., N$ and is still Gaussian. So, we integrate over $\psi, \bar\psi$ to obtain
\begin{equation}\label{eq:Oma}
    \Omega(\alpha) = c_N^{(BD)} \!\!\!\! \int\limits_{M_0 \times M_1} \!\!\!\! D\mu(Z,\widetilde{Z}) \, \mathrm{SDet}^{N/2}(1 - \widetilde{Z} Z) \, \mathrm{SDet}^{-N/2} \begin{pmatrix} \alpha &Z \cr \widetilde{Z} &\alpha \end{pmatrix} .
\end{equation}

To process this expression further, it is useful to think of the $n \times n$ supermatrices $Z, \widetilde{Z}$ as off-diagonal blocks in a bigger matrix and perform a Cauchy transform:
\begin{equation}
    \begin{pmatrix} 0 &Z \cr \widetilde{Z} &0 \end{pmatrix}
    = \frac{Q\Sigma_3 - 1}{Q\Sigma_3 + 1} \equiv \Gamma_Q \,,
\end{equation}
introducing $Q = T \Sigma_3 T^{-1}$ with $\Sigma_3 = \mathrm{diag}(1_n \,, -1_n)$ as the supermatrix that runs through an adjoint orbit $\mathrm{Ad} (G) \cdot \Sigma_3 \cong G/K$ parametrized (locally) by $T \in G$,
\begin{equation}\label{eq:rat-T}
    T = \begin{pmatrix} 1 &Z \cr \widetilde{Z} &1 \end{pmatrix} \begin{pmatrix} (1-Z\widetilde{Z})^{-1/2} &0 \cr 0 &(1-\widetilde{Z} Z)^{-1/2} \end{pmatrix} ,
\end{equation}
where $G$ is a suitable real form of the complex Lie supergroup $\widetilde{\mathrm{OSp}}(2n_0 \vert 2n_1)$ over $\mathrm{Sp}(2n_0,\mathbb{C}) \times \mathrm{SO}(2n_1,\mathbb{C})$. The intermediate result (\ref{eq:Oma}) then takes the final form
\begin{equation}\label{eq:Oma-fin}
    \Omega(\alpha) = \int\limits_{M_0 \times M_1} \!\!\!\! DQ \; \mathrm{SDet}^{N/2} \left( \frac{1 +\Gamma_Q}{\alpha_2 + \Gamma_Q} \right) , \quad \alpha_2 = \mathrm{diag}(\alpha,\alpha) ,
\end{equation}
with $DQ \equiv c_N^{(BD)} D\mu (Z,\widetilde{Z})$ the (normalized) $G$-invariant Berezin integral form on $G/K$.

In the limit of large $N$, the integral (\ref{eq:Oma-fin}) can be done by the saddle-point approximation (which is actually exact here, by the Duistermaat-Heckman principle \cite{DH} of semiclassical exactness). The saddle points are given by
diagonal matrices
\begin{equation}
    Q = \mathrm{diag}(1_{n_0},s; - 1_{n_0},-s) , \quad s = \mathrm{diag} (s_1, \ldots, s_{n_1}) ,
\end{equation}
where the $s_\nu$ take values in $\{\pm 1\}$ and are subject to the constraint $\sum s_\nu \in n_1 - 4 \mathbb{Z}$. The result for $\Omega(\alpha)$ is a sum over all saddle points, parameterized by $s$ subject to the stated constraint. Without loss of generality, we may specialize to the case of $n_0 = n_1 = n/2$. Then
\begin{equation}\label{eq:answer}
    \Omega(\alpha) =  \prod_{\mu \leq n_0} \alpha_\mu^{-N} \sum_s F_s(\alpha) \prod_{n_0 < \nu} \alpha_\nu^{N(1+s_\nu)/2} \,,
\end{equation}
where the factor $F_s(\alpha)$ accounts for the Gaussian fluctuations around each saddle point:
\begin{equation}
    F_s(\alpha) = \frac{\prod_{1 \leq \mu \leq n_0 < \nu \leq n} (1-\alpha_\mu^{-1} \alpha_\nu^{-s_\nu})}{\prod_{\mu \leq \mu^\prime \leq n_0} (1-\alpha_\mu^{-1} \alpha_{\mu^\prime}^{-1}) \prod_{n_0 < \nu < \nu^\prime} (1-\alpha_\nu^{-s_\nu} \alpha_{\nu^\prime}^{-s_{\nu^\prime}})} \,.
\end{equation}

Now we come to the main message of this subsection. In \cite{HPZ}, it was proved by different techniques that the result (\ref{eq:answer}) holds exactly not just for large $N$ but for \underline{all} positive integers $N$. Thus, the present calculation using the CF transformation happens to give the correct result for all $N$, and there is no indication of a breakdown of the CF transformation for small $N$. The absence of any such indication might be interpreted as a signal that the CF transformation for the supersymmetric case ($n_0 = n_1$) could be correct for all $N \geq 1$. Alas, that interpretation turns out to be too optimistic: while the specific result (\ref{eq:answer}) is protected by certain miraculous phenomena due to supersymmetry, the same phenomena do not protect the CF transformation against instability for small $N$, as we shall see.

\subsection{Color-flavor transformation of type $C$}

The CF transformation of type $C$ was already presented in the original reference \cite{suprev}. Its setup is basically the same as for type $BD$ (Section \ref{sect:2.1}), so we shall be very brief here, describing only what changes from before.

Assuming $N \in 2\mathbb{N}$ we replace the Euclidean structure $\delta$ of $\mathbb{R}^N$ by a symplectic structure $\varepsilon$ with tensor components
\begin{equation}
    \varepsilon_{ij} = - \varepsilon_{ji} \,, \quad \varepsilon^{ij} \varepsilon_{jk} = \delta^i_k \,.
\end{equation}
(The symplectic structure $\varepsilon$ must be compatible with the Hermitian scalar product on the Hilbert space $\mathbb{C}^N \supset \mathbb{R}^N$, i.e., if $\psi^i$ are the components of a unit vector, then the same must be true for $\varepsilon_{ij} \psi^j$.) The unitary symplectic group $\mathrm{Sp}(N) \equiv \mathrm{USp}(N)$ is defined by the relations $g^{-1} = g^\dagger$ and
\begin{equation}
    g_{\;k}^i \varepsilon_{ij} g_{\;l}^j = \varepsilon_{kl} \quad \big( g \in \mathrm{Sp}(N) \big).
\end{equation}
It is a basic fact of invariant theory \cite{Howe-Schur} that all invariant tensors of $\mathrm{Sp}(N)$ arise at the quadratic level, i.e.\ are polynomials in the degree-2 invariant $\varepsilon_{ij}$. The basic integrals over $\mathrm{Sp}(N)$ are
\begin{equation}
    \int\limits_{\mathrm{Sp}(N)}\!\!\! dg = 1 , \quad \int\limits_{\mathrm{Sp}(N)} \!\!\! dg\; g_{\;j}^i = 0 , \quad \int\limits_{\mathrm{Sp}(N)} \!\!\! dg\; g_{\;k}^i g_{\;l}^j = N^{-1} \varepsilon^{ij} \varepsilon_{lk} \,.
\end{equation}
The replacement of the symmetric form $\delta_{ij}$ by the skew-symmetric form $\varepsilon_{ij}$ modifies the exchange symmetries of the square supermatrices $Z$ and $\widetilde{Z}$:
\begin{equation}
    Z^{\mu\nu} = -(-1)^{|\mu| |\nu| + |\mu| + |\nu|} Z^{\nu\mu} , \quad  \widetilde{Z}_{\mu\nu} = - (-1)^{|\mu||\nu|} \widetilde{Z}_{\nu\mu} \,.
\end{equation}
The statement \cite{circular} of the CF transformation of type $C$ now is that
\begin{eqnarray}\label{eq:type-C}
    &&\int_{\mathrm{Sp}(N)} \!\!\!\! dg \, \exp \left(  g_{\; j}^i \bar\psi_{\; \mu}^j \psi_{\; i}^\mu \right) = c_N^{(C)} \int_{M_0 \times M_1} \!\!\!\!\!\! D\mu (Z,\widetilde{Z}) \, \mathrm{SDet}^{N/2}(1 - \widetilde{Z} Z) \\ &&\hspace{4cm} \times \exp \left( {\textstyle{\frac{1}{2}}} \varepsilon^{ik} \psi_{\;k}^\nu  (-1)^{|\nu|} \widetilde{Z}_{\nu\mu} \psi_{\;i}^\mu + {\textstyle{\frac{1}{2}}} \bar\psi_{\;\mu}^j Z^{\mu\nu} \bar\psi_{\;\nu}^l \varepsilon_{lj} \right) \nonumber
\end{eqnarray}
holds with a finite constant $c_N^{(C)}$ when the number of colors $N$ satisfies $N \geq 2n_0-2$. Here, the expression for the $G$-invariant Berezin integral form changes to
\begin{equation}\label{eq:Dm-typeC}
    D\mu(Z,\widetilde{Z}) = D(Z,\widetilde{Z}) \circ \mathrm{SDet}^{-n_0 + n_1 + 1}(1 - \widetilde{Z} Z) ,
\end{equation}
while the normalization constant $c_N^{(C)}$ is still determined by the analog of Eq.\ (\ref{eq:cN}), with $D\mu(Z,\widetilde{Z})$ given by Eq.\ (\ref{eq:Dm-typeC}). The Hermitian symmetric spaces $M_0$ and $M_1$ making up the integration domain on the right-hand side are now $M_1 = \mathrm{Sp}(2n_1) / \mathrm{U}(n_1)$ and $M_0 = \mathrm{SO}^\ast(2n_0) / \mathrm{U}(n_0)$, the latter being a noncompact form of $\mathrm{SO} (2n_0) / \mathrm{U}(n_0)$.

\subsection{Color-flavor transformation of type $A$}\label{sect:A-type}

As usual, the terminology ``type $A$'' refers to the family of unitary groups, $\mathrm{U}(N)$. That family differs from the orthogonal and symplectic families in that the non-trivial part of any polynomial in the matrix elements $g_{\; j}^i$ of $g \in \mathrm{U}(N)$ integrates to zero against the Haar measure $dg :$
\begin{equation}
    \int\limits_{\mathrm{U}(N)}\!\!\!\! dg = 1 , \quad \int\limits_{\mathrm{U}(N)} \!\!\!\! dg\; g_{\;j}^i = 0 , \quad \int\limits_{\mathrm{U}(N)} \!\!\!\! dg\; g_{\;j}^i \, g_{\;l}^k = 0 , \quad {\rm etc.},
\end{equation}
due to the absence of $\mathrm{U}(N)$-invariants in tensor powers of the defining representation space $\mathbb{C}^N$ for $\mathrm{U}(N)$. Nonzero invariants are found in tensor products of $\mathbb{C}^N$ with the dual representation space, $(\mathbb{C}^N)^\ast$. Again, from invariant theory \cite{Howe-Schur} one knows that all invariants arise at the quadratic level, where one has just a single invariant, namely the tautological invariant $\mathrm{Id} \in \mathrm{End}(\mathbb{C}^N) \cong \mathbb{C}^N \otimes (\mathbb{C}^N)^\ast$ -- or in components: $\delta_j^i\,$. The basic integral is
\begin{equation}
    \int\limits_{\mathrm{U}(N)} \!\!\!\! dg \; g_{\;j}^i (g^{-1})_{\;l}^k = N^{-1} \delta_l^i \delta_j^k .
\end{equation}
The type-$A$ situation with inequivalent representations $\mathbb{C}^N$ and $(\mathbb{C}^N)^\ast$ calls for an expanded setup with ``advanced'' and ``retarded'' integration variables, as follows.

The variables of the advanced sector are denoted by $\psi_{\;i}^\mu$ with color index $i = 1, \ldots, N$ and flavor index $\mu = 1, \ldots, m$. As before, they are commuting for $|\mu| = 0$ and anti-commuting for $|\mu| = 1$. There are $m_0$ commuting and $m_1$ anti-commuting flavors, summing up to $m_0 + m_1 = m$. The conjugate variables $\bar\psi_{\;\mu}^{i}$ have the same index range and fermion parity as the variables $\psi_{\;i}^\mu \,$. All of these are used to form color-space quadratic arrays $\bar\psi_{\;\mu}^i \psi_{\;j}^{\mu}\,$, which are coupled to unitary $N \times N$ matrices $g$ as
\begin{equation}
    (g^{-1})_{\;i}^j \bar\psi_{\;\mu}^i \psi_{\;j}^{\mu} \,.
\end{equation}

The variables of the retarded sector are denoted by $\varphi^{\nu i}$ and $\bar\varphi_{i \nu}$. The color index still is $i = 1, \ldots, N$, but the range of the flavor index ($\nu = 1, \ldots, n = n_0 + n_1$) may differ from that of the advanced sector. Flavor invariants made from retarded variables couple to color-group elements in the vector representation:
\begin{equation}
    \bar\varphi_{i\nu} g_{\ j}^i \varphi^{\nu j} .
\end{equation}

We now introduce in flavor space a rectangular complex supermatrix $Z$ of size $m \times n$ with matrix elements $Z^{\mu\nu}$ and another such matrix $\widetilde{Z}$ of size $n \times m$ with matrix elements $\widetilde{Z}_{\nu\mu}\,$. The even-even subblock of $\widetilde{Z}$ (of size $n_0 \times m_0$) is taken to be the Hermitian adjoint of the corresponding $m_0 \times n_0$ subblock of $Z$; the odd-odd subblocks of $\widetilde{Z}$ and $Z$ (of sizes $n_1 \times m_1$ resp.\ $m_1 \times n_1$) are anti-Hermitian adjoints of each other. Together with the range restriction (\ref{eq:restr}), which remains in force, these relations ensure that the numerical parts of the supermatrices $1 - \widetilde{Z} Z$ and $1 - Z \widetilde{Z}$ are always positive.

In that setting, the CF transformation of type $A$ takes the form
\begin{eqnarray}\label{eq:type-A}
    &&\int_{\mathrm{U}(N)} \!\!\!\! dg \, \exp \left(  (g^{-1})_{\;i}^j \bar\psi_{\;\mu}^i \psi_{\;j}^{\mu} + g_{\;j}^i  \varphi^{\nu j} (-1)^{|\nu|} \bar\varphi_{i\nu} \right) \\ &&= c_N^{(A)} \int_{M_0 \times M_1}\!\!\!\!\!\! D\mu (Z,\widetilde{Z}) \, \mathrm{SDet}^{N}(1 - \widetilde{Z} Z) \exp \left( \varphi^{\nu j} (-1)^{|\nu|} \widetilde{Z}_{\nu\mu} \psi_{\;j}^{\mu} + \bar\psi_{\;\mu}^i Z^{\mu\nu} \bar\varphi_{i\nu} \right) . \nonumber
\end{eqnarray}
We expect it to hold with finite $c_N^{(A)}$ when $N$ is large enough. The normalization constant $c_N^{(A)}$ is determined by
\begin{equation}\label{eq:cNA}
    c_N^{(A)} \int_{M_0 \times M_1} \!\!\!\! D\mu (Z,\widetilde{Z}) \, \mathrm{SDet}^N (1-\widetilde{Z} Z) = 1 .
\end{equation}
The factors of the domain $M_0 \times M_1$ of integration are symmetric spaces,
\begin{eqnarray}
    &&M_0 = \mathrm{U}(m_0,n_0) / \mathrm{U}(m_0) \times \mathrm{U}(n_0) , \cr
    &&M_1 = \mathrm{U}(m_1 + n_1) / \mathrm{U}(m_1) \times \mathrm{U}(n_1) ,
\end{eqnarray}
known as complex Grassmann manifolds of noncompact and compact type.

Actually, in order to avoid complications that arise in the super-setting (e.g., in cases like $m_0 = n_1 > m_1 = n_0\,$, where no maximal torus of even type exists), we will restrict our attention here and in the following to the case of $m_0 = n_0$ and $m_1 = n_1$ (i.e., equal numbers of retarded and advanced flavors). In that case, the supermatrices $Z$ and $\widetilde{Z}$ remain square, and the 
condition of validity of the color-flavor transformation (\ref{eq:type-A}) is $N \geq 2n_0\,$. Also, the $G$-invariant Berezin integral form still has a simple expression:
\begin{equation}\label{eq:Dm-typeA}
    D\mu(Z,\widetilde{Z}) = D(Z,\widetilde{Z}) \circ \mathrm{SDet}^{-2n_0 + 2n_1}(1 - \widetilde{Z} Z) .
\end{equation}

\section{Mathematical origin of the color-flavor transformation} \label{sect:origin}
\setcounter{equation}{0}

Our text so far has repeatedly made reference to various basic facts from the invariant theory of the classical groups $\mathrm{U}(N)$, $\mathrm{O}(N)$, and $\mathrm{Sp}(N)$. In the current section, we are going to explain that our whole concept of color-flavor transformation is firmly rooted in invariant theory. More precisely, every CF transformation described above is a corollary to a fundamental principle known as \emph{Howe duality}. A brief overview is as follows.

Howe duality makes a statement about so-called \emph{Howe pairs} acting reductively on a Fock space for either bosons or fermions \cite{Howe-Schur} or for both particle types \cite{Howe-Trans}. In our structured setting with supersymmetry, a Howe pair is understood to be a pair $(U,\mathfrak{g})$ with the following properties: (i) $U$ is a compact classical Lie group; (ii) $\mathfrak{g}$ is a complex Lie superalgebra; (iii) the actions (or representations) of $U$ and $\mathfrak{g}$ on Fock space are reductive (i.e.\ decompose into irreducibles) and commute with each other and (iv) the latter property is maximal (one also says that $U$ and $\mathfrak{g}$ are mutual centralizers). There exist three types of such pairs $(U,\mathfrak{g})$:
\begin{equation}
    \begin{array}{ll} \mathrm{U}(N), &\mathfrak{gl}(n_0 \vert n_1) = \mathfrak{gl}(n_0) \oplus \mathfrak{gl}(n_1) \oplus \ldots \,, \\ \mathrm{Sp}(N), &\mathfrak{osp}(2n_0 \vert 2n_1) = \mathfrak{o}(2n_0) \oplus \mathfrak{sp}(2n_1) \oplus \ldots \, , \cr \mathrm{O}(N), &\widetilde{\mathfrak{osp}}(2n_0 \vert 2n_1) = \mathfrak{sp}(2n_0) \oplus \mathfrak{o}(2n_1) \oplus \ldots \,,
    \end{array} \nonumber
\end{equation}
where the ellipses indicate the odd part of the Lie superalgebra. In each case, the action of $U$ on Fock space preserves the particle number, while $\mathfrak{g}$ includes generators that act as pair creation and pair annihilation operators.

The key assertion of Howe duality is that the reductive Howe pair $(U,\mathfrak{g})$ acts on Fock space \emph{without multiplicity}, i.e., each irreducible (joint) representation of the pair $(U,\mathfrak{g})$ occurs at most once. In particular, the multiplicity space of the trivial $U$-representation (a.k.a.\ the space of color-neutral states) is an irreducible representation space for the Howe partner $\mathfrak{g}$ of $U$. The latter fact directly leads to the CF transformation of type $A$, $C$, and $BD$, respectively. To provide some detail, we will focus on the third pair -- the one that gives rise to the CF transformation of type $BD$.

\subsection{Restating the CF transformation (type $BD$)}\label{sect:derive}

Since Howe duality is a statement about the joint action of a Howe pair $(U,\mathfrak{g})$ on Fock space, we now turn to the requisite formalism, which is that of second quantization. Recalling our notational conventions for color ($i = 1, \ldots, N$) and flavor ($\mu = 1, \ldots, n$), we introduce single-particle annihilation operators $a_{\ i}^{\,\mu}$ and the corresponding creation operators $\bar{a}_{\;\mu}^{\,i}$; they are bosonic for $|\mu| = 0$ and fermionic for $|\mu| = 1$. As usual, the Fock vacuum $| 0 \rangle$ is defined by demanding that
\begin{equation}
    a_{\ i}^{\,\mu} | 0 \rangle = 0 \quad (i=1,\ldots,N; \; \mu=1,\ldots,n) .
\end{equation}
The canonical (anti-)commutation relations are succinctly written as
\begin{equation}\label{eq:CAR}
    [ a_{\ i}^{\,\mu} \,, \bar{a}_{\; \nu}^{\,j} ] \equiv a_{\ i}^{\,\mu} \bar{a}_{\; \nu}^{\,j} - (-1)^{|\mu| |\nu|} \bar{a}_{\; \nu}^{\,j} a_{\ i}^{\,\mu} = \delta_i^j \delta_\nu^\mu ,
\end{equation}
and
\begin{equation}\label{eq:CCR}
    [ a_{\ i}^{\,\mu} \,, a_{\;j}^{\,\nu} ] = 0 = [ \bar{a}_{\;\mu}^{\,i} \,, \bar{a}_{\;\nu}^{\,j} ] .
\end{equation}
Note that the superbracket $[ \cdot \,, \cdot ]$ is the commutator unless both single-particle operators are fermionic (i.e., $|\mu| = |\nu| = 1$), in which case it is the anti-commutator.

Next, we recall our integration variables $\psi, \bar\psi$ and give them a new interpretation -- we regard them as variables that parameterize coherent states (for bosons and fermions):
\begin{equation}\label{eq:costates}
    \langle \psi \vert = \langle 0 \vert \exp \left( \bar\psi_{\;\mu}^{\,i} a_{\ i}^{\,\mu} \right) , \quad \vert \psi \rangle = \exp \left( \bar{a}_{\;\mu}^{\,i} \psi_{\ i}^{\mu} \right) \vert 0 \rangle .
\end{equation}
Here (and throughout), the Koszul sign rule is in force: $\bar\psi_{\;\mu}^{\,i} a_{\ i}^{\,\mu} = (-1)^{|\mu|} a_{\ i}^{\,\mu} \bar\psi_{\;\mu}^{\,i} \,$, etc. For present use, we note that the overlap between boson/fermion coherent states is
\begin{equation}
    \langle \psi \vert \psi^\prime \rangle = \exp \left( \bar\psi_{\;\mu}^{\,i} {\psi^\prime}_{\ i}^{\mu} \right) .
\end{equation}

Both the color group $U = \mathrm{O}(N)$ and the flavor algebra $\mathfrak{g} = \widetilde{\mathfrak{osp}}(2n_0 | 2n_1)$ have representations by operators that act on the Fock space of states generated from the vacuum $| 0 \rangle$ by the repeated application of the $\bar{c}_{\;\mu}^{\,i}\,$. The $\mathfrak{g}$-representation will be described in the next subsection. For present purposes, we just invoke the $\mathrm{O}(N)$-representation (without writing it down explicitly), which is given by a map of second quantization, $g \mapsto \mathcal{D}(g)$, with the property that
\begin{equation}
    \mathcal{D}(g) \bar{a}_{\;\mu}^{\,i} \mathcal{D}(g)^{-1} = (g^{-1})_{\;j}^i \bar{a}_{\;\mu}^{\,j} \,.
\end{equation}
It follows that
\begin{equation}
    \langle \psi \vert \mathcal{D}(g) \vert \psi^\prime \rangle = \exp \left( (g^{-1})_{\; j}^i \bar\psi_{\;\mu}^j {\psi^\prime}_{\;i}^\mu \right) .
\end{equation}
[Here is the reason why we prefer $g^{-1}$, instead of $g$, in Eq.\ (\ref{eq:ginv}).] On basic grounds, the normalized Haar integral of the representation $\mathcal{D}(g)$,
\begin{equation}\label{eq:intro-P}
    \int_{\mathrm{O}(N)} \!\!\! dg \; \mathcal{D}(g) =
    \int_{\mathrm{O}(N)} \!\!\! dg \; \mathcal{D}(g^{-1}) \equiv P ,
\end{equation}
has the effect of orthogonal projection, $P$, onto the subspace of color-neutral states, carrying the trivial representation of $\mathrm{O}(N)$. (One also speaks of color-singlet states.) Thus the left-hand side of the CF transformation (\ref{eq:type-D}) can be rewritten as a coherent-state expectation of the color-singlet projector $P$:
\begin{equation}\label{eq:LHS-CFT}
    \int_{\mathrm{O}(N)} \!\!\! dg\; \exp \left(  g_{\; j}^i \bar\psi_{\; \mu}^j {\psi^\prime}_{\;i}^\mu \right) = \langle \psi \vert P \vert \psi^\prime \rangle \,.
\end{equation}
The CF transformation (\ref{eq:type-D}) now comes about because there exists an alternative way of expressing $P$, by employing the Howe partner $\mathfrak{g} = \widetilde{\mathfrak{osp}} (2n_0 | 2n_1)$ which is assigned to $U = \mathrm{O}(N)$ by Howe duality. This needs some explaining, as follows.

Acting on the Fock space, the Lie superalgebra $\mathfrak{g}$ is generated by the operators
\begin{equation}\label{eq:LSA-osp}
    a_{\ i}^{\,\mu} \delta^{ij} a_{\ j}^{\,\nu} \,, \quad \bar{a}_{\;\mu}^{\,i} \delta_{ij} \bar{a}_{\;\nu}^{\,j} \,, \quad {\textstyle{\frac{1}{2}}} \left( a_{\ i}^{\,\mu} \bar{a}_{\;\nu}^{\,j} + (-1)^{|\mu||\nu|} \bar{a}_{\;\nu}^{\,j} a_{\ i}^{\,\mu} \right) .
\end{equation}
Here the particle-number conserving operators, $\bar{a} a$ and $a \bar{a}$, are ``Weyl-ordered'': they are sym\-metrized in the boson-boson and skew-symmetrized in the fermion-fermion sector.

The Fock vacuum $\vert 0 \rangle$, a color-neutral state, may serve as a highest-weight vector for the irreducible $\mathfrak{g}$-representation which is furnished by the collection of color-neutral states. Simply put, any color-neutral state can be obtained by repeatedly applying operators from the pair-creation set  $\{\bar{a}_{\;\mu}^{\,i} \delta_{ij} \bar{a}_{\;\nu}^{\,j} \}$ to the Fock vacuum. Hence, following a standard procedure, we build unnormalized ``spin''-coherent states by coherent pair creation on top of the vacuum:
\begin{eqnarray}
    &&\vert Z \rangle = \exp\Big( {\textstyle{\frac{1}{2}}} \bar{a}_{\;\mu}^{\,i} Z^{\mu\nu} \bar{a}_{\;\nu}^{\,j} \delta_{ji} \Big) \vert 0 \rangle , \label{eq:3.11} \\ &&\langle Z \vert = \langle 0 \vert \exp\left( {\textstyle{\frac{1}{2}}} \delta^{ij} a_{\ j}^{\,\nu} (-1)^{|\nu|} \widetilde{Z}_{\nu\mu} a_{\ i}^{\,\mu} \right) . \label{eq:3.12}
\end{eqnarray}
Taking overlaps with the coherent states (\ref{eq:costates}), we find
\begin{equation}
    \langle \psi \vert Z\rangle \langle Z\vert \psi\rangle = \exp \left( {\textstyle{\frac{1}{2}}} \bar\psi_{\;\mu}^i Z^{\mu\nu} \bar\psi_{\;\nu}^j \delta_{ji} + {\textstyle{\frac{1}{2}}} \delta^{ij} \psi_{\;j}^\nu (-1)^{|\nu|} \widetilde{Z}_{\nu\mu} \psi_{\;i}^\mu \right) .
\end{equation}
Now the overlap between unnormalized spin-coherent states is
\begin{equation}
    \langle Z \vert Z^\prime \rangle = \mathrm{SDet}^{-N/2} (1 - \widetilde{Z} Z^\prime ) ,
\end{equation}
and by comparing the expressions above with Eq.\ (\ref{eq:type-D}), we see that our CF trans\-formation can be restated as
\begin{equation}\label{eq:3.16}
    \langle \psi \vert P \vert \psi^\prime \rangle = c_N^{(BD)} \int D\mu(Z,\widetilde{Z}) \, \mathrm{SDet}^{N/2}(1 - \widetilde{Z} Z) \,  \langle \psi \vert Z\rangle \langle Z\vert \psi^\prime \rangle \,.
\end{equation}
We know that the boson/fermion coherent states $\vert \psi \rangle$ form an (overcomplete) basis. Therefore, in view of the reformulation (\ref{eq:3.16}), proving the statement (\ref{eq:type-D}) of the CF transformation is equivalent to showing that we have
\begin{equation}
    c_N^{(BD)} \int D\mu(Z,\widetilde{Z}) \, \mathrm{SDet}^{N/2}(1 - \widetilde{Z} Z) \;  \vert Z\rangle \langle Z\vert = P .
\end{equation}
In words: the spin-coherent states $| Z \rangle$ need to furnish a resolution of the identity $P$ on the subspace of color-neutral states. More concisely, using $\mathrm{SDet}^{ -N/2} (1-\widetilde{Z} Z) = \langle Z | Z \rangle$, the relation at stake is
\begin{equation}\label{eq:key-rel}
    P \stackrel{?}{=} c_N^{(BD)} \int D\mu (Z,\widetilde{Z}) \; \vert Z\rangle \langle Z | Z \rangle^{-1} \langle Z\vert \,.
\end{equation}
Our strategy for proving the relation (\ref{eq:key-rel}) is laid out in Section \ref{sect:strategy}.

\subsection{Fock representation of $\widetilde{\mathfrak{osp}}$}\label{sect:Fock-rep}

For the sequel, we need the concrete form of the Fock-space representation, say $X \mapsto \widehat{X}$, of $\mathfrak{g} = \widetilde{\mathfrak{osp}}(2n_0|2n_1)$. To describe it, we start with the remark that $\mathfrak{g}$ is $\mathbb{Z}$-graded by a special element $\Sigma_3 \in \mathfrak{g} :$
\begin{equation}
    \Sigma_3 = \mathbf{1}_{n_0 \vert n_1} \otimes \sigma_3 \,,
\end{equation}
which might be called the ``particle number'' as it acts on the Fock space as the Weyl-ordered particle-number operator:
\begin{equation}\label{eq:Sigma3}
    \widehat{\Sigma_3} = {\textstyle{\frac{1}{2}}} \left( \bar{a}_{\;\mu}^{\,i} a_{\ i}^{\,\mu}
    + (-1)^{|\mu|} a_{\ i}^{\,\mu}\, \bar{a}_{\;\mu}^{\,i} \right) .
\end{equation}
The adjoint (or commutator) action of $\Sigma_3$ on the Lie superalgebra $\mathfrak{g}$ decomposes the latter into three subalgebras:
\begin{equation}
    \mathfrak{g} = \mathfrak{n}_- \oplus \mathfrak{h} \oplus \mathfrak{n}_+ \,,
\end{equation}
which are spanned by the elements with root (or eigenvalue) $-2$, $0$, $+2$, and are represented by pair annihilation, particle-number conserving, and pair creation operators, respectively. We adopt a matrix representation in which
\begin{equation}
    \begin{pmatrix} 0 &B \cr 0 &0 \end{pmatrix} \in \mathfrak{n}_+ \,, \quad
    \begin{pmatrix} 0 &0 \cr C &0 \end{pmatrix} \in \mathfrak{n}_- \,,
\end{equation}
and
\begin{equation}
    \begin{pmatrix} A &0 \cr 0 &-A^{\rm sT} \end{pmatrix} \in \mathfrak{h} .
\end{equation}
Here $A^{\rm sT}$ is a supertranspose of $A$; in components: $(A^{\rm sT})_{\nu}^{\ \mu} = (-1)^{|\mu|(|\nu|+1)} A_{\ \nu}^\mu \,$.

Next, to take care of the sign issues beleaguering super-representations, we pass from $\mathfrak{g}$ to a ``Grassmann envelope'' thereof (Berezin \cite{Berezin}); i.e., we help ourselves to some parameter Grassmann algebra $\Lambda = \Lambda^{\rm even} \oplus \Lambda^{\rm odd}$ and fill the even matrix entries of the blocks $A$, $B$, $C$ with elements from $\Lambda^{\rm even}$, and the odd matrix entries with elements from $\Lambda^{\rm odd}$. By that process, $B$ is turned into a supermatrix of the type of our $Z$ above, and $C$ becomes like $\widetilde{Z}$. Please be warned that to avoid overloading our notation, we will \underline{not} decorate the symbols to indicate the passage to a Grassmann envelope but assume that step to be understood.

Given these conventions, we can specify the Fock-space representation $X \mapsto \widehat{X}$. Schematically speaking, the representation is formed by sandwiching the supermatrix $X$ between single-particle creation and annihilation operators:
\begin{equation}
    X = \begin{pmatrix} A &B \cr C &-A^{\rm sT} \end{pmatrix} \mapsto \widehat{X} =  \frac{1}{2} \Big( \bar{a} \quad \mp a \Big) \begin{pmatrix} A &B \cr C &-A^{\rm sT} \end{pmatrix} \begin{pmatrix} a \cr \bar{a} \end{pmatrix} .
\end{equation}
Speaking precisely, the representation is expressed as
\begin{eqnarray}\label{eq:g-rep}
    \widehat{X} &=& {\textstyle{\frac{1}{2}}} \bar{a}_{\;\mu}^{\,i} A_{\;\nu}^\mu a_{\ i}^{\,\nu}
    + {\textstyle{\frac{1}{2}}} \bar{a}_{\;\mu}^{\,i} B^{\mu\nu} \bar{a}_{\;\nu}^{\,j} \delta_{ji} \cr &+& {\textstyle{\frac{1}{2}}}  \delta^{ij} a_{\ j}^{\,\nu} (-1)^{|\nu|+1} C_{\nu\mu} a_{\ i}^{\,\mu} + {\textstyle{\frac{1}{2}}} a_{\ i}^{\,\nu} (-1)^{|\nu|+1} (- A^{\rm sT})_{\nu}^{\ \mu} \bar{a}_{\;\mu}^{\,i} \,,
\end{eqnarray}
where we recognize the $B$- and $C$-summands as the terms that already appeared in the exponents of Eqs.\ (\ref{eq:3.11}, \ref{eq:3.12}). We remark that the sign factors $(-1)^{|\nu|}$ in Eq.\ (\ref{eq:g-rep}) appear because the fundamental superbracket (\ref{eq:CAR}) behaves under exchange as
\begin{equation}
    [ a_{\; i}^{\nu} \,, \bar{a}_{\;\mu}^{\,i} ] =
    [ \bar{a}_{\;\mu}^{\,i} \,, a_{\; i}^{\nu} ] (-1)^{|\nu|} .
\end{equation}
Now, using the (anti-)commutation relations (\ref{eq:CAR}, \ref{eq:CCR}) (along with the Koszul sign rule), it is not difficult to verify that the correspondence $X \mapsto \widehat{X}$ is indeed a Lie algebra isomorphism (or representation).

\subsection{Strategy of proof}\label{sect:strategy}

With the $\mathfrak{g}$-representation in hand, we turn to our proof strategy for the relation (\ref{eq:3.16}) or, equivalently, (\ref{eq:key-rel}). First of all, it must be stressed that the integral on the right-hand side of, say, Eq.\ (\ref{eq:3.16}) does not exist in the whole range of parameters $N$, $n_0$, $n_1\,$. Roughly speaking, a failure occurs when the number $N$ of colors lies below a certain threshold. Thus, to prove the color-flavor transformation, an essential step is to ascertain the existence of the integral. That step will be taken in Section \ref{sect:bounds}; in the present section, we take the existence of the integral for granted; i.e., we assume that $N$ is large enough.

To begin the discussion, let $G_\mathbb{C}$ be the complex Lie supergroup which is obtained by exponentiating the complex Lie superalgebra $\mathfrak{g} = \widetilde{\mathfrak{osp}}(2n_0|2n_1)$ in its (super-)matrix representation. In the original article \cite{circular}, there was a tacit assumption that the $\mathfrak{g}$-representation $X \mapsto \widehat{X}$ exponentiates to a $G_\mathbb{C}$-representation, $g \equiv \mathrm{e}^X \mapsto \mathcal{D}(g) = \mathrm{e}^{\widehat{X}}$, on Fock space. Alas, that assumption is not mathematically tenable. To indicate the nature of the problem, let $H > 0$ be a harmonic-oscillator Hamiltonian (acting on Hilbert space). While $H$ does generate a \emph{semigroup} $\beta \mapsto \mathrm{e}^{-\beta H}$ ($\mathrm{Re}\, \beta > 0$), it does \underline{not} exponentiate to a \emph{group}, as $\mathrm{e}^{- \beta H}$ does not exist as an operator on Hilbert space when $\mathrm{Re} \, \beta < 0$. In the present case, the situation is similar. What can be exponentiated are the elements of a half-space in $\mathfrak{g}\,$, giving rise to a semigroup of contractions in $G_\mathbb{C}\,$.

Thus we are facing the inconvenience that a group representation $g \mapsto \mathcal{D}(g)$ does not really exist, and the original argument made in Ref.\ \cite{circular} needs to be modified. In principle, one does know how to get around the complication: the case of the harmonic-oscillator semigroup was worked out in detail by Howe in \cite{Howe-oscrep}, and the relevant mathematics was adapted to the supercase in Refs.\ \cite{CFZ, HPZ}. However, in the present paper we wish to keep the reasoning as simple as possible. Therefore we shall avoid $G_\mathbb{C}$ for now and build our argument around the $\mathfrak{g}$-representation (\ref{eq:g-rep}), which is well-defined without further ado.

The argument begins with the obvious statement that the color-singlet projector $P$ acts as the identity on the space of color-singlet states. Since our operators $\widehat{X}$ transform color singlets into other color singlets, we immediately have
\begin{equation}\label{eq:comm-rel}
    \widehat{X} P - P \widehat{X} = 0 \quad ({\rm for \; all \;} X \in \Lambda^{\rm even} \otimes \widetilde{\mathfrak{osp}}_{\rm even} \oplus \Lambda^{\rm odd} \otimes \widetilde{\mathfrak{osp}}_{\rm odd}) .
\end{equation}
Now recall the Howe duality statement that the $\mathfrak{g}$-representation on color-neutral states is irreducible. By Schur's Lemma, it follows that the vector space of solutions to the commutation relation (\ref{eq:comm-rel}) is complex one-dimensional and hence given by the scalar multiples of $P$. (Please be assured that while the standard proof of Schur's Lemma assumes a finite-dimensional irreducible representation, it is easily adapted \cite{HPZ} to the present case of an infinite-dimensional irrep with the Fock vacuum as highest-weight vector). Thus if we can show that the right-hand side of (\ref{eq:key-rel}) is another solution to the commutation relation (\ref{eq:comm-rel}), we will have proved that the two sides of (\ref{eq:key-rel}) are the same but for a scalar factor (which we absorb into the normalization constant $c_N^{(BD)}$).

To summarize: taking the existence of integrals for granted, our task of proving the CF transformation (\ref{eq:type-D}) has been reduced to showing that the right-hand side of (\ref{eq:key-rel}) commutes with all operators $\widehat{X}$.
We take a closer look at that task in the next subsection.

\subsection{From commutators to derivations}

To establish the desired property of the right-hand side of (\ref{eq:key-rel}), we have to work out the commutators of the pure-state density operator
\begin{equation}\label{eq:def-rho}
    \rho(Z,\widetilde{Z}) \equiv | Z \rangle \langle Z \vert Z \rangle^{-1} \langle Z \vert
\end{equation}
with all the operators $\widehat{X}$. We will now do so for the individual parts of $\widehat{X}$ as given in Eq.\ (\ref{eq:g-rep}), starting with block-type $B$ (from $\mathfrak{n}_+ \subset \mathfrak{g}$). Let
\begin{equation}
    {\textstyle{\frac{1}{2}}} \bar{a}_{\;\mu}^{\,i} B^{\mu\nu} \bar{a}_{\;\nu}^{\,j} \delta_{ji} \equiv \widehat{B}
\end{equation}
for short. Then, recalling Eq.\ (\ref{eq:3.11}), we immediately see that
\begin{equation}
    \widehat{B} \, \vert Z \rangle = {\sum}^\prime B^{\mu\nu} \frac{\partial}{\partial Z^{\mu\nu}} \vert Z \rangle \,,
\end{equation}
where ${\sum}^\prime$ means that the sum is restricted to run over a set of pairs $(\mu\nu)$ for which the matrix elements $Z^{\mu\nu}$ of the supermatrix $Z$ with exchange symmetry (\ref{eq:exch-Z}) are \emph{independent} variables. With a little more effort, we deduce from Eq.\ (\ref{eq:3.12}) that
\begin{eqnarray}
   \langle Z \vert\, \widehat{B} &=& \langle 0 \vert\, \mathrm{e}^{\frac{1}{2} \delta^{ij} a_{\ j}^{\,\nu} (-1)^{|\nu|} \widetilde{Z}_{\nu\mu} a_{\ i}^{\,\mu} } \widehat{B} \cr &=& \bigg( {\sum}^\prime (\widetilde{Z} B \widetilde{Z})_{\mu\nu} \frac{\partial}{\partial \widetilde{Z}_{\mu\nu}} + \frac{N}{2} \mathrm{STr}\, \widetilde{Z} B \bigg) \langle Z \vert \,;
\end{eqnarray}
the calculational strategy here is simply this: commute $\widehat{B}$ to the left of the exponential (where it annihilates the Riesz-dual $\langle 0 \vert$ of the Fock vacuum) and collect the bracket terms that are produced along the way.

Next, we observe that
\begin{equation}\label{eq:3.32}
    \frac{N}{2} \mathrm{STr}\, \widetilde{Z} B = {\sum}^\prime \left( B^{\mu\nu} \frac{\partial}{\partial Z^{\mu\nu}} - (\widetilde{Z} B \widetilde{Z})_{\mu\nu} \frac{\partial}{\partial \widetilde{Z}_{\mu\nu}} \right) \ln \langle Z \vert Z \rangle .
\end{equation}
Finally, taking matrix elements between any two states and putting together all the relations above, we obtain
\begin{equation}\label{eq:cover-B}
    \langle \psi \vert [ \widehat{B} , \rho(Z,\widetilde{Z}) ] \vert \psi^\prime \rangle = - \delta_B \, \langle \psi \vert \rho(Z,\widetilde{Z}) \vert \psi^\prime \rangle
\end{equation}
with first-order differential operator (or derivation)
\begin{equation}
    \delta_B = - {\sum}^\prime B^{\mu\nu} \frac{\partial}{\partial Z^{\mu\nu}} + {\sum}^\prime (\widetilde{Z} B \widetilde{Z})_{\mu\nu} \frac{\partial}{\partial \widetilde{Z}_{\mu\nu}} .
\end{equation}
To stream-line the notation, let us abbreviate this as
\begin{equation}\label{eq:Kill-B}
    \delta_B = - \Big\langle B , \partial_Z \Big\rangle + \Big\langle \widetilde{Z} B \widetilde{Z}, \partial_{\widetilde{Z}} \Big\rangle .
\end{equation}

Second, we consider the operators $\widehat{X}$ of block-type $C$ (from $\mathfrak{n}_- \subset \mathfrak{g}$):
\begin{equation}
    - {\textstyle{\frac{1}{2}}}  \delta^{ij} c_{\ j}^{\,\nu} (-1)^{|\nu|} C_{\nu\mu} c_{\ i}^{\,\mu} \equiv \widehat{C} .
\end{equation}
Going through the same calculational steps as before, we find
\begin{equation}\label{eq:cover-C}
    \langle \psi \vert [ \widehat{C} , \rho(Z,\widetilde{Z}) ] \vert \psi^\prime \rangle = - \delta_C \, \langle \psi \vert \rho(Z,\widetilde{Z}) \vert \psi^\prime \rangle
\end{equation}
where
\begin{equation}\label{eq:Kill-C}
    \delta_C = \Big\langle Z C Z , \partial_Z \Big\rangle - \Big\langle C , \partial_{\widetilde{Z}} \Big\rangle .
\end{equation}
Third, we turn to the operators $\widehat{X}$ of block-type $A$ (from $\mathfrak{h} \subset \mathfrak{g}$):
\begin{equation}
    {\textstyle{\frac{1}{2}}} \bar{a}_{\;\mu}^{\,i} A_{\;\nu}^\mu a_{\ i}^{\,\nu} + {\textstyle{\frac{1}{2}}} a_{\ i}^{\,\nu} (-1)^{|\nu|} (A^{\rm sT})_{\nu}^{\ \mu} \bar{a}_{\;\mu}^{\,i} \equiv \widehat{A} =  \bar{a}_{\;\mu}^{\,i} A_{\;\nu}^\mu a_{\ i}^{\,\nu} + {\rm const} .
\end{equation}
In this case we obtain
\begin{equation}\label{eq:cover-A}
    \langle \psi \vert [ \widehat{A} , \rho(Z,\widetilde{Z}) ] \vert \psi^\prime \rangle = - \delta_A \, \langle \psi \vert \rho(Z,\widetilde{Z}) \vert \psi^\prime \rangle
\end{equation}
with
\begin{equation}\label{eq:Kill-A}
    \delta_A = - \Big\langle A Z + Z A^{\rm sT}, \partial_Z \Big\rangle + \Big\langle \widetilde{Z} A + A^{\rm sT} \widetilde{Z} , \partial_{\widetilde{Z}} \Big\rangle .
\end{equation}
In summary, all matrix elements $\langle \psi \vert [ \widehat{X} , \rho(Z,\widetilde{Z}) ] \vert \psi^\prime \rangle$ for $X$ from $\mathfrak{g} = \widetilde{\mathfrak{osp}}(2n_0|2n_1)$ can be expressed as some first-order differential operator applied to the function $\langle \psi \vert \rho(Z,\widetilde{Z}) \vert \psi^\prime \rangle$. In the decomposition $X = C + A + B$ according to $\mathfrak{g} = \mathfrak{n}_- \oplus \mathfrak{h} \oplus \mathfrak{n}_+\,$, the corresponding first-order differential operator $\delta_C + \delta_A + \delta_B$ is given by Eqs.\ (\ref{eq:Kill-C}, \ref{eq:Kill-A}, \ref{eq:Kill-B}).

A mathematical summary of the situation is this: the $\langle \psi \vert Z \rangle$ transform w.r.t.\ the given $\mathfrak{g}$-action as holomorphic sections of a $\mathbb{C}$-line bundle, the $\langle Z \vert \psi^\prime \rangle$ transform as anti-holomorphic sections, and the $\langle \psi \vert \rho(Z,\widetilde{Z}) \vert \psi^\prime \rangle$ transform as functions on the underlying Hermitian symmetric superspace of Cartan-type $C{\rm I}\vert D{\rm I\!I\!I}$ with local coordinates given by the matrix entries of the supermatrices $Z$ and $\widetilde{Z}$.

\subsection{The derivations $\delta_\bullet$ are Killing vector fields}\label{sect:killing}

We now enter the next stage of our proof strategy for the color-flavor trans\-formation. The upshot of Section \ref{sect:strategy} was that the type-$BD$ identity (\ref{eq:type-D}) holds true if we can show that the right-hand side of Eq.\ (\ref{eq:key-rel}) commutes with all operators $\widehat{X}$ for $X$ from $\mathfrak{g} = \widetilde{\mathfrak{osp}} (2n_0|2n_1)$. In the preceding subsection, we expressed all such commutators (actually, their matrix elements between any two states) as first-order differential operators $\delta_\bullet$ acting on the density matrix elements $\langle \psi \vert \rho(Z,\widetilde{Z}) \vert \psi^\prime \rangle$ viewed as functions of $Z$ and $\widetilde{Z}$. Our next step is to recognize these derivations $\delta_\bullet$ as \emph{total} derivatives. With that step accomplished, we can hope to carry out the integral
\begin{equation}\label{eq:vanish}
    \int D\mu(Z,\widetilde{Z}) \; \delta_\bullet \langle \psi \vert \rho(Z,\widetilde{Z}) \vert \psi^\prime \rangle \stackrel{?}{=} 0
\end{equation}
and thus establish the desired commutation property, leading to the desired outcome.

To prepare the following, we pause for a brief tutorial on differentiation and integration in the super-context. Consider $\mathbb{R}^{p| q}$ with even coordinate functions $x^1 , \ldots, x^p$ and odd coordinates (a.k.a.\ Grassmann variables) $x^{p+1}, \ldots, x^{p+q}$. Let
\begin{equation}
    {D}x \equiv dx^1 \cdots dx^p \, \frac{\partial}{\partial x^{p+1}} \cdots
    \frac{\partial}{\partial x^{p+q}}
\end{equation}
be a flat Berezin integral form. Then if $F$ is a compactly supported superfunction for $\mathbb{R}^{p|q}$, all of its total derivatives (even or odd) integrate to zero:
\begin{equation}\label{eq:pI0}
    \int_{\mathbb{R}^p} {D}x \, \frac{\partial}{\partial x^\alpha} F = 0 \quad (\alpha = 1, \ldots, p+q) .
\end{equation}
Next let ${D}\mu = {D}x \circ \Omega$ be a non-flat Berezin integral form deformed by a superfunction $\Omega = \Omega(x)$, and let
\begin{equation}
    \delta_V = V^\alpha (x) \frac{\partial}{\partial x^\alpha}
\end{equation}
be a first-order differential operator. Without loss for our purposes, we may assume that both $\Omega$ and $\delta_V$ are even and that the numerical value of $\Omega(x)$ is invertible for all points in the domain $\mathbb{R}^p$. (Note that in the case of $\delta_V$ we have to allow for a dependence of the coefficients $V^\alpha$ on extraneous Grassmann parameters due to our Grassmann envelope construction for $\mathfrak{g}\,$, so the adjective ``even'' here is meant to take into account the Koszul sign rule). Then the formal adjoint of $\delta_V$ with respect to $D\mu$ is expressed as
\begin{equation}
    \delta_V^\ast = - (-1)^{|\alpha|} \frac{1}{\Omega} \frac{\partial}{\partial x^\alpha} \circ \Omega V^\alpha .
\end{equation}
Indeed, for any superfunction $E$ which is even (a choice made in order to avoid the appearance of some trivial signs) one has
\begin{equation}\label{eq:pI1}
    \int D\mu \, E \delta_V F = \int D\mu \, (\delta_V^\ast E) F + \int Dx \, \frac{\partial}{\partial x^\alpha} \left( \Omega V^\alpha E F \right) (-1)^{|\alpha|},
\end{equation}
so $\delta_V^\ast$ is adjoint to $\delta_V$ up to an integral of total derivatives, which vanishes according to Eq.\ (\ref{eq:pI0}) if $EF$ is compactly supported.

A special class of derivations is constituted by so-called \emph{Killing vector fields}, which have the characteristic property of being their own formal skew-adjoints:
\begin{equation}\label{eq:kill}
    \delta_V^\ast = - \delta_V .
\end{equation}
For such a derivation, the partial integration (\ref{eq:pI1}) with $E \equiv 1$ simplifies to
\begin{equation}\label{eq:pI2}
    \int D\mu \, \delta_V F = \int Dx \, \frac{\partial}{\partial x^\alpha} \left( \Omega V^\alpha F \right) (-1)^{|\alpha|}
\end{equation}
due to $\delta_V^\ast E = - \delta_V 1 = 0$. Thus for a differential operator $\delta_V$ with the Killing property (\ref{eq:kill}), the derivative $\delta_V F$ is a total derivative ready to be integrated. To conclude this short tutorial, we observe that (\ref{eq:kill}) is equivalent to the requirement that $\delta_V$ has vanishing super-divergence with respect to $D\mu = Dx \circ \Omega:$
\begin{equation}\label{eq:sdiv}
    \mathrm{SDiv}_{D\mu} ( \delta_V ) \equiv (-1)^{|\alpha|} \frac{1}{\Omega} \frac{\partial}{\partial x^\alpha} \left(\Omega V^\alpha \right)  = 0 .
\end{equation}

We now apply the general principles above to the specific case at hand. We recall from Eqs.\ (\ref{eq:invt-Dm}, \ref{eq:flat-D}) that our Berezin integral form $D\mu(Z,\widetilde{Z})$ is the flat form $D(Z,\widetilde{Z})$ preceded by multiplication with the function
\begin{equation}
    \Omega \equiv \mathrm{SDet}^{- n_0 + n_1 - 1}(1 - \widetilde{Z} Z) .
\end{equation}
For that choice of Berezin integral form, we now compute the super-divergence (\ref{eq:sdiv}) of the first-order differential operators $\delta_A$, $\delta_B$, $\delta_C$ of Eqs.\ (\ref{eq:Kill-A}, \ref{eq:Kill-B}, \ref{eq:Kill-C}). Starting again with $\delta_B$, we observe that
\begin{equation}\label{eq:3.48}
    \delta_B \ln\Omega = - (n_0 - n_1 + 1)\, \mathrm{STr}\, \widetilde{Z} B .
\end{equation}
(This is none other than Eq.\ (\ref{eq:3.32}) with $N/2$ replaced by $n_0 - n_1 + 1$.) Now according to  the definition (\ref{eq:sdiv}), we have
\begin{equation}\label{eq:3.49}
    \mathrm{SDiv}_{D\mu}(\delta_B) = \delta_B \ln\Omega + {\sum}^\prime (-1)^{|\mu| + |\nu|} \frac{\partial}{\partial \widetilde{Z}_{\mu\nu}} (\widetilde{Z} B \widetilde{Z})_{\mu\nu} \,.
\end{equation}
To calculate the restricted sum ${\Sigma}^\prime$ correctly, we break it up into three pieces: the boson-boson sector (where $Z$ is symmetric), the fermion-fermion sector (where $Z$ is skew), and the boson-fermion mixed sector (where half of the matrix entries of $Z$ are dependent on the other half). The respective terms due to differentiation of the left factor $\widetilde{Z}$ in $(\widetilde{Z} B \widetilde{Z})_{\mu\nu}$ are
\begin{equation}
    \left( \frac{n_0+1}{2} \mathrm{Tr}_{\rm even} + \frac{n_1-1}{2} \mathrm{Tr}_{\rm odd} - \frac{n_0}{2} \mathrm{Tr}_{\rm odd} - \frac{n_1}{2} \mathrm{Tr}_{\rm even} \right) (B \widetilde{Z}) ;
\end{equation}
they add up to $\frac{1}{2} (n_0 - n_1 + 1)\, \mathrm{STr} \, B \widetilde{Z}$. Differentiation of the right factor in $(\widetilde{Z} B \widetilde{Z})_{\mu\nu}$ results in the same contribution. Thus, when the contributions from both factors, left and right, are taken into account, Eq.\ (\ref{eq:3.49}) becomes
\begin{equation}
    \mathrm{SDiv}_{D\mu}(\delta_B) = \delta_B \ln\Omega + (n_0 - n_1 + 1)\, \mathrm{STr} \, \widetilde{Z} B .
\end{equation}
In view of (\ref{eq:3.48}), this implies that $\mathrm{SDiv}_{D\mu}(\delta_B) = 0$. Thus $\delta_B$ has vanishing super-divergence w.r.t.\ $D\mu$ and therefore satisfies the Killing property (\ref{eq:kill}) entailing (\ref{eq:pI2}). This is no surprise, as $D\mu$ is the Berezin integral form given by the supergeometry (\ref{eq:inv-metric}) with symmetry group $G$, and $\delta_B$ (as a derivation due to the Lie superalgebra $\mathfrak{g}$ of $G$) was constructed by (the infinitesimal version of) the same symmetry principle.

Turning from $\delta_B$ to the derivations $\delta_A$ and $\delta_C$ of Eqs.\ (\ref{eq:Kill-A}, \ref{eq:Kill-C}) we remark that these, too, are Killing vector fields, by the very same symmetry principle. The concrete calculations for them are not very different, and the final result is easily checked to remain the same:
\begin{equation}\label{eq:ABC-kill}
    \mathrm{SDiv}_{D\mu}(\delta_A) = \mathrm{SDiv}_{D\mu}(\delta_B) = \mathrm{SDiv}_{D\mu}(\delta_C) = 0 .
\end{equation}
In words: all our first-order differential operators $\delta_\bullet$ (for $\bullet = A$, $B$, or $C$) have vanishing s-divergence with respect to $D\mu$. It follows that they are formally skew-self-adjoint [Eq.\ (\ref{eq:kill})] and lend themselves to integration as total derivatives [Eq.\ (\ref{eq:pI2})].

Summarizing all the mathematical reasoning compounded in Sections \ref{sect:derive}--\ref{sect:killing}, we arrive at the following statement.

\medskip\noindent\textbf{Fact 1.} The color-flavor transformation of type $BD$ holds true, as formulated in Eq.\ (\ref{eq:type-D}), and with a finite normalization constant $c_N^{(BD)}$ determined by Eq.\ (\ref{eq:cN}), if the integral on the right-hand side of the former equation exists.
 
\medskip\noindent\textit{Proof.} What remains to be shown is that all integrals (\ref{eq:vanish}) vanish. Since our first-order differential operators $\delta_\bullet$ for $\bullet = A, B, C$ all have the Killing property (\ref{eq:ABC-kill}), the integrands of (\ref{eq:vanish}) are total derivatives. The vanishing of the integrals is then immediate by inspection of the expressions (\ref{eq:Kill-B}), (\ref{eq:Kill-C}), and (\ref{eq:Kill-A}).

 
\subsection{Special cases}\label{sect:special}

Fact 1, stated at the end of the preceding subsection, gives a precise criterion by which to decide whether the identity (\ref{eq:type-D}) of the color-flavor transformation holds true or not. Let us now put that criterion to work in a few easy cases.

\medskip\noindent i. Fermions only ($n_0 = 1$). In that case we have
\begin{equation}
    \Omega = \mathrm{Det}^{-(n_1-1)} \big( 1 - \widetilde{Z}_{\rm FF} Z^{\rm FF} \big) , \quad \widetilde{Z}_{\rm FF} = - \big( Z^{\rm FF} \big)^\dagger ,
\end{equation}
and the measure $D\mu$ is an $\mathrm{SO}(2n_1)$-invariant measure on the type-$D{\rm I\!I\!I}$ symmetric space $M_1 \cong \mathrm{SO}(2n_1) / \mathrm{U}(n_1)$ with finite volume $\int_{M_1} D\mu < \infty$. (Note that $M_1$ for $n_1 = 1$ consists of just a single point. This triviality is a simple consequence of the $\mathrm{O}(N)$-invariant bilinear form on $\mathbb{R}^N$ being symmetric and of Fermi statistics.) The integrand on the right-hand side of Eq.\ (\ref{eq:type-D}) is a matrix element of the density operator $\rho(Z,\widetilde{Z})$ in Eq.\ (\ref{eq:def-rho}); as such it is a continuous function on the compact space $M_1$ and therefore remains bounded for any matrix element(s) of $Z^{\rm FF}$ going to infinity. Hence the integral on the r.h.s.\ of (\ref{eq:type-D}) is guaranteed to exist.
%
%
It follows that the CF transformation (\ref{eq:type-D}) for $n_0 = 0$ holds for all $n_1 \geq 1$ and all $N \geq 1$, without restriction.

\medskip\noindent ii. Bosons only ($n_1 = 0$). Now
\begin{equation}\label{eq:Om-BD-bos}
    \Omega = \mathrm{Det}^{-(n_0 + 1)} \big( 1 - \widetilde{Z}_{\rm BB} Z^{\rm BB} \big) , \quad \widetilde{Z}_{\rm BB} = + \big( Z^{\rm BB} \big)^\dagger ,
\end{equation}
and $D\mu$ is a $G$-invariant measure on $M_0 \cong G/K \equiv \mathrm{Sp}(2n_0,\mathbb{R}) / \mathrm{U}(n_0)$, a noncompact symmetric space of type $C{\rm I}$, here parameterized by the complex symmetric matrix $Z^{\rm BB}$ with noncompact domain $(Z^{\rm BB})^\dagger Z^{\rm BB} < 1$. That space has infinite volume. Therefore, the range of the parameters $n_0$ and $N$ must be restricted in order for the integral (\ref{eq:type-D}) to converge. What helps toward convergence is the presence of
\begin{equation}
    \langle Z \vert Z \rangle^{-1} = \mathrm{Det}^{N/2}\big( 1 - \widetilde{Z}_{\rm BB} Z^{\rm BB} \big) .
\end{equation}
For $N \geq 2n_0+1$, the decrease of that factor offsets enough of the growth of $\Omega$ on approaching the boundary of $M_0$ at infinity (where one or several singular values of the complex matrix $Z^{\rm BB}$ go to unity), thereby making the integral on the r.h.s.\ of Eq.\ (\ref{eq:type-D}) converge.

One might think that an even larger value of $N$ would be needed in order for the integrals (\ref{eq:vanish}) to vanish, but that is not so. Indeed, for the basic case of $n_0 = 1$, where $Z^{\rm BB} \equiv z$ reduces to a single complex variable with domain $|z| < 1$, we can do directly the following calculation. The Killing vector fields $\delta_\bullet$ (of block-type $\bullet = A, B, C$) are
\begin{eqnarray}
    &&\delta_{\tiny \left( \begin{array}{cc} \mathrm{i} &0 \cr 0 &-\mathrm{i} \end{array} \right)}
    = - 2 \mathrm{i}\, ( z \partial_z - \bar{z} \partial_{\bar{z}} ) , \cr
    &&\delta_{\tiny \left( \begin{array}{cc} 0 &1 \cr 0 &0 \end{array} \right)} = - \partial_z + \bar{z}^2 \partial_{\bar{z}} \,, \quad \delta_{\tiny \left( \begin{array}{cc} 0 &0 \cr 1 &0 \end{array} \right)} = - \partial_{\bar{z}} + z^2 \partial_z \,.
\end{eqnarray}
Let us consider the block-type $B$ vector field with formal adjoint
\begin{equation}
    - \delta_{\tiny \left( \begin{array}{cc} 0 &1 \cr 0 &0 \end{array} \right)} =
    \delta^\ast_{\tiny \left( \begin{array}{cc} 0 &1 \cr 0 &0 \end{array} \right)} =
    \Omega^{-1} ( \partial_z - \partial_{\bar{z}} \, \bar{z}^2 )\, \Omega
\end{equation}
and total-derivative integral
\begin{eqnarray}\label{eq:3.63}
    \int_{|z| < 1} D\mu\; \delta_{\tiny \left( \begin{array}{cc} 0 &1 \cr 0 &0 \end{array} \right)} F &\propto& \int dz \wedge d\bar{z}\, ( \partial_z - \partial_{\bar{z}} \, \bar{z}^2 )\, \Omega F \cr &=& \int d \left( \Omega F ( d\bar{z} + \bar{z}^2 dz ) \right) .
\end{eqnarray}
By Stokes' Theorem, the last expression gets integrated to a line integral along the boundary circle $|z| = 1$, where the one-form in the integrand vanishes identically:
\begin{equation}
    d\bar{z} + \bar{z}^2 dz \Big\vert_{z = \mathrm{e}^{\mathrm{i} \theta}} = - \mathrm{i}\, \mathrm{e}^{- \mathrm{i}\theta} d\theta + \mathrm{i}\, \mathrm{e}^{-\mathrm{i}\theta} d\theta = 0 .
\end{equation}
In order for the integral (\ref{eq:3.63}) to vanish, the one-form $d\bar{z} + \bar{z}^2 dz$ multiplied by the differentiable function $\Omega F$ must still go to zero (on approaching the boundary $|z| = 1$). That is the case for $N \geq 3$ (and $n_0 = 1$). The situation and outcome for the block-type $A$ and $C$ Killing vector fields $\delta_\bullet$ is similar. We thus conclude that the bosonic (i.e., $n_1 = 0$) CF transformation (\ref{eq:type-D}) holds true for all $N \geq 3$, in the special case of $n_0 = 1$.

For higher rank, $n_0 > 1$, one can do an analogous calculation, but there actually exists a more elegant way to proceed. In the development of Section \ref{sect:strategy}, we shied away from working with the $G$-action (to avoid the technical issues of exponentiating a Lie-algebra half-space to a semigroup action). However, in the present instance all the necessary mathematics is in place and available. The relevant semigroup is known as the oscillator semigroup \cite{Howe-oscrep}, and its closure houses the so-called metaplectic group $\mathrm{Mp}(2n_0)$ [a double cover of $\mathrm{Sp}(2n_0,\mathbb{R})$], acting by unitary operators on the bosonic Fock space. If $\widehat{T} \mapsto T$ is the covering map $\mathrm{Mp} \to \mathrm{Sp}$, and $(Z,\widetilde{Z}) \mapsto (T \cdot Z, T\cdot \widetilde{Z})$ expresses the action of $G = \mathrm{Sp}(2n_0, \mathbb{R})$ on $M_0 \cong G/K$ in the coordinate matrices $Z \equiv Z^{\rm BB}$, $\widetilde{Z} \equiv (Z^{\rm BB})^\dagger$, one has the relation
\begin{equation}
    \int D\mu \, \langle \psi \vert \widehat{T} \rho(Z,\widetilde{Z}) \widehat{T}^{\,-1} \vert \psi^\prime \rangle = \int D\mu \, \langle \psi \vert \rho(T^{-1} \cdot Z, T^{-1} \cdot \widetilde{Z}) \vert \psi^\prime \rangle ,
\end{equation}
combining and strengthening the infinitesimal relations (\ref{eq:cover-B}, \ref{eq:cover-C}, \ref{eq:cover-A}) of before. The symplectic group $G$ acts on the noncompact integration domain $Z \widetilde{Z} < 1$. Hence, after a substitution of integration variables, $T^{-1} \cdot Z = Z^\prime$ and $T^{-1} \cdot \widetilde{Z} = \widetilde{Z}^\prime$, our whole argument based on Schur's Lemma and $G$-invariance boils down to checking whether $D\mu$ with density function $\Omega$ given by Eq.\ (\ref{eq:Om-BD-bos}) is $G$-invariant. That invariance property is easily confirmed. In conclusion, the condition of existence of the integral on the r.h.s.\ of Eq.\ (\ref{eq:type-D}) is sufficient, and the CF transformation of bosonic type ($n_1 = 0$) holds for all $N \geq 2 n_0 + 1$. In representation theory \cite{stable}, that range is sometimes called the ``stable'' range for the holomorphic discrete series of $\mathrm{Mp}(2n_0)$.

Now in Ref.\ \cite{FK}, the bosonic CF transformation of type $A$ has been extended beyond the stable range. One might therefore ask whether a similar extension is possible here. That question will be addressed in Section \ref{sect:3.8}, albeit only in the simple case of $n_0 = 1$.

\medskip\noindent iii. Boson-Fermion mixed case, $n_0 = n_1 = n$. This case is of considerable interest for applications, say to $\mathrm{O}(N)$-network models of disordered superconductors. Since the addition of compact degrees of freedom due to the fermionic variables is unlikely to aggravate the situation with respect to convergence, we expect that the CF still holds for $N \geq 2 n_0 + 1$ (see Section \ref{sect:bounds} for the precise argument). That range excludes a number of small values of $N$ desirable for model building. Unfortunately, it seems to be no easy job to figure out whether (and if so, how) the threshold for $N$ can be lowered. We will make a first step in that direction in the next subsection.

\subsection{Reproducing kernel}

In the sequel, we shall address the question as to whether the range of validity of the CF transformation can be extended to lower $N$-values, possibly by modifying the detailed expression of the transformation. We begin with some mathematical background.

We recall that the noncompact integration domain $M_0$ given by $0 \leq Z^{\rm BB} (Z^{\rm BB})^\dagger < 1$ has a boundary at infinity, which is reached by one or several singular values of the complex matrix $Z^{\rm BB}$ tending to unity. The main issue here is that this boundary domain houses a number of distinct $G$-invariant orbits of the action of the group $G$, and the number of such orbits increases with $n_0$. [For the case of the bosonic type-$BD$ CF transformation and for $n_0 \in 2\mathbb{N}$, the $G$-orbit of smallest dimension, given as the solution space of $Z^{\rm BB} (Z^{\rm BB})^\dagger = 1$, is a symmetric space $\mathrm{U}(n_0) / \mathrm{Sp}(n_0)$, also called the Shilov-boundary of $M_0\,$.] The proliferation of $G$-invariant orbits entails a proliferation of $G$-invariant integrals. Therefore, when the boundary of the noncompact domain $M_0$ comes into play (due to integrands decreasing slowly for small $N$), the right-hand side of the color-flavor transformation becomes non-unique (if one uses nothing but symmetry arguments), if it does not fail altogether.

To facilitate our further discussion and enable generalizations, we recall the key expression (\ref{eq:key-rel}) for the color-singlet projector $P = P^2$, and we reformulate it as follows. We take matrix elements between two spin-coherent states on both sides. Then, since $P \vert Z \rangle = \vert Z \rangle$, Eq.\ (\ref{eq:key-rel}) turns into
\begin{equation}
    \langle Z^\prime \vert Z^{\prime\prime} \rangle \stackrel{?}{=} c_N^{(BD)}
    \int D\mu (Z, \widetilde{Z}) \; \langle Z^\prime \vert Z \rangle \, \langle Z \vert Z \rangle^{-1} \langle Z\vert Z^{\prime\prime} \rangle .
\end{equation}
{}From this we read off the following easy-to-use criterion: the CF transformation (\ref{eq:type-D}) holds true if and only if the overlap kernel (a.k.a.\ Bergman kernel)
\begin{equation}
    K_N (\widetilde{Z},Z^\prime) \equiv \mathrm{SDet}^{-N/2} (1 - \widetilde{Z} Z^\prime) = \langle Z \vert Z^\prime \rangle
\end{equation}
reproduces under integration with Berezin integral form $c_N^{(BD)} D\mu (Z, \widetilde{Z}) \circ \langle Z \vert Z \rangle^{-1}$:
\begin{equation}\label{eq:easy-crit}
    K_N(\widetilde{Z}^\prime \vert Z^{\prime\prime}) \stackrel{?}{=} c_N^{(BD)}
    \int D\mu (Z, \widetilde{Z}) \, \langle Z \vert Z \rangle^{-1} K_N( \widetilde{Z}^\prime ,Z)
    K_N(\widetilde{Z}, Z^{\prime\prime}) .
\end{equation}

For the color-flavor transformation of type $C$ one has the same criterion, but with $D\mu(Z,\widetilde{Z})$ given by Eq.\ (\ref{eq:Dm-typeC}) and the normalization constant $c_N^{(BD)}$ replaced by $c_N^{(C)}$. For the color-flavor transformation of type $A$ one still has the same criterion, but with $D\mu(Z,\widetilde{Z})$ given by Eq.\ (\ref{eq:Dm-typeA}) and another normalization constant $c_N^{(A)}$. Note that the type-$A$ overlap kernel is
\begin{equation}
    K_N^{(A)} (\widetilde{Z},Z^\prime) = \mathrm{SDet}^{-N} (1 - \widetilde{Z} Z^\prime) .
\end{equation}

\subsection{Type $BD$, $n_1=0$, $n_0=1$, $N$ small}\label{sect:3.8}

We now revisit the bosonic type-$BD$ situation of a color group $\mathrm{O}(N)$, with $n_1 = 0$ (no fermionic flavors) and $n_0 = 1$ (one bosonic flavor), and focusing on small $N$. In that setting, the Howe partner of $U = \mathrm{O}(N)$ is the Lie algebra $\mathfrak{g} = \mathfrak{sp}(2, \mathbb{R})$ of the noncompact Lie group $G = \mathrm{Sp}(2,\mathbb{R})$. In keeping with the general conventions of the present paper, we continue to work in the coordinate-picture of a complex matrix $Z$, which here is just a complex number $Z \equiv z$, with complex conjugate $\widetilde{Z} = \bar{z}$. The use of complex coordinates (for the complex manifold $M_0$) lets us see the symplectic Lie group and its Lie algebra by their incarnations as the right-hand sides of the accidental isomorphisms
\begin{equation}
    G = \mathrm{Sp}(2,\mathbb{R}) \cong \mathrm{SU}(1,1) , \quad \mathfrak{g} = \mathfrak{sp}(2,\mathbb{R}) \cong \mathfrak{su}(1,1) ,
\end{equation}
which are assumed to be understood. The noncompact integration domain defined by $|z| < 1$ is also known as the Poincar\'e disk model of the two-dimensional hyperboloid
\begin{equation}
    M_0 = \mathrm{H}^2 \cong G/K, \quad K = \mathrm{SO}(2) \cong \mathrm{U}(1) .
\end{equation}
The noncompact Lie group $G \cong \mathrm{SU}(1,1)$ has a parametrization by two complex numbers $\alpha$ and $\beta$ subject to the hyperbolic constraint $|\alpha|^2 - |\beta|^2 = 1$. In the coordinates $z$ and $\bar{z}$, the group $G$ acts on the hyperboloid $G/K$ by rational transformations
\begin{equation}\label{eq:Sp2-param}
    \mathrm{SU}(1,1) \ni g = \begin{pmatrix} \alpha &\bar\beta \cr \beta &\bar\alpha \end{pmatrix} : \; z \mapsto g \cdot z = \frac{\alpha z + \bar\beta}{\beta z+ \bar\alpha} \,, \quad g \cdot \bar{z} = \overline{g \cdot z} \,,
\end{equation}
and the $G$-invariant integration measure on $M_0 = \mathrm{H}^2$ is
\begin{equation}
    D\mu(Z,\widetilde{Z}) = (1-|z|^2)^{-2} d^2 z ,
\end{equation}
with Lebesgue measure $d^2 z$ normalized by $\int_{|z| < 1} d^2 z = \pi$.

We now ask whether the overlap kernel for the case of $N$ colors,
\begin{equation}
    K_N(\bar{z},z^\prime) = (1 - \bar{z} z^\prime)^{-N/2} ,
\end{equation}
reproduces under convolution with integration measure
\begin{equation}
    c_N^{(BD)} D\mu(Z,\widetilde{Z}) \langle Z | Z \rangle^{-1} = c_N^{(BD)} (1-|z|^2)^{N/2-2} d^2 z .
\end{equation}
As always, the normalization constant $c_N^{(BD)}$ is fixed by the demand (\ref{eq:cN}) for a total mass of unity, which yields
\begin{equation}
    c_N^{(BD)} = \frac{N-2}{2\pi}\,.
\end{equation}
Thus the reproducing property (\ref{eq:easy-crit}) required of the overlap kernel $K_N(\bar{z},z^\prime)$ reads
\begin{equation}\label{eq:testBD}
    K_N(\bar{z}^\prime,z^{\prime\prime}) = \frac{N-2}{2\pi} \int_{|z| < 1} \!\!\!\! d^2 z\, (1-|z|^2)^{N/2-2} K_N(\bar{z}^\prime,z) K_N (\bar{z}, z^{\prime\prime}) .
\end{equation}
To check the validity of the identity (\ref{eq:testBD}) we expand $K_N$ as a power series:
\begin{equation}\label{eq:pow-ser}
    K_N(\bar{z},z^\prime) = \frac{1}{\Gamma(N/2)} \sum_{m=0}^\infty \frac{\Gamma(m+N/2)}{\Gamma(m+1)} (\bar{z} z^\prime)^m
\end{equation}
and do the integral over $|z| < 1$ term by term of the sum over $m$:
\begin{equation}
    \frac{N-2}{2\pi} \int_{|z| < 1} \!\!\!\! d^2 z\, (1-|z|^2)^{N/2-2} |z|^{2m} = \frac{\Gamma(N/2) \Gamma(m+1)}{\Gamma(m+N/2)} .
\end{equation}
We then see that Eq.\ (\ref{eq:testBD}) is fine as long as $N \geq 3$, confirming what we saw earlier.


However, there does exist a problem for $N = 1$ and $N = 2$. In the latter case, $c_{N=2}^{(BD)}$ vanishes and, concomitantly, the integral (\ref{eq:testBD}) diverges. It turns out that this $0 \times \infty$ dilemma can be resolved, as follows. The open Poincar\'e disk $|z| < 1$ is an orbit for the action (\ref{eq:Sp2-param}) of $\mathrm{SU}(1,1) \cong \mathrm{Sp}(2,\mathbb{R})$, but so is the boundary circle $\mathrm{S}^1 :\; |z| = 1$. Indeed, using the parametrization in (\ref{eq:Sp2-param}) one has
\begin{equation}
    \mathrm{S}^1 \ni \mathrm{e}^{\mathrm{i}\theta} \mapsto g \cdot \mathrm{e}^{\mathrm{i}\theta} = \frac{\alpha\, \mathrm{e}^{ \mathrm{i} \theta} + \bar\beta}{\beta\, \mathrm{e}^{\mathrm{i}\theta} + \bar\alpha} \in \mathrm{S}^1 .
\end{equation}
This observation spawns the idea to move all of the ``mass'' of the $G$-invariant integral from the two-dimensional Poincar\'e disk $\mathrm{H}^2 = G/K$ to the one-dimensional boundary circle $\mathrm{S}^1$. To implement that idea, we parameterize the boundary circle $|z| = 1$ by $z = \mathrm{e}^{\mathrm{i}\theta}$ and observe that the reproducing property
\begin{equation}
    K_2(\mathrm{e}^{-\mathrm{i}\theta^\prime}, \mathrm{e}^{\mathrm{i}\theta^{\prime\prime}})
    = \int_0^{2\pi} \frac{d\theta}{2\pi} \; K_2(\mathrm{e}^{-\mathrm{i}\theta^\prime} , \mathrm{e}^{\mathrm{i} \theta}) K_2(\mathrm{e}^{-\mathrm{i}\theta} , \mathrm{e}^{\mathrm{i} \theta^{\prime\prime}}),
\end{equation}
with adapted integration measure $d\theta / 2\pi$, does hold for the overlap kernel $K_{N=2}$ restricted to the boundary circle:
\begin{equation}
    K_2(\mathrm{e}^{-\mathrm{i}\theta}, \mathrm{e}^{\mathrm{i}\theta^{\prime}}) = (1-\mathrm{e}^{-\mathrm{i}(\theta - \theta^{\prime})})^{-1} ,
\end{equation}
at least when applied to the boundary limits $f(z = \mathrm{e}^{\mathrm{i}\theta})$ of holomorphic functions $f(z)$ taken from Hardy space (to ensure convergence). Thus, a resolution of the $\mathrm{O}(N)$-singlet projector $P$ by spin-coherent states still exists even for the problematic case of $N = 2$. By consequence, the type-$BD$ color-flavor transformation (\ref{eq:type-D}) with $n_0 = 1$ and $n_1 = 0$ extends down to $N = 2$, albeit in modified form (with a modified domain $M_0 = \mathrm{S}^1$ and with adapted integration measure and normalization constant for $N=2$).

What about $N = 1$? There, the color-neutral sector of the bosonic Fock space is none other than the (parity-even sector of the) Hilbert space of the one-dimensional harmonic oscillator. In that case, no reproducing kernel built on spin-coherent states (as opposed to boson-coherent states) exists, and our color-flavor transformation fails, with no possibility for repair. A sure indication of the failure is that the (analytically continued) normalization constant $c_{N=1}^{(BD)}$ of a (positive) measure becomes negative. Looking more closely, we see from Eq.\ (\ref{eq:pow-ser}) that one would need a probability measure on the unit disk $|z| \leq 1$ with moments
\begin{equation}
    \mathbb{E} \left( |z|^{2m} \right) = \frac{\Gamma(1/2) \Gamma(m+1)}{\Gamma(m+1/2)} \,,
\end{equation}
which increase with $m$ even though $|z|^{2m}$ (on the disk) does not increase with $m$. Clearly, such a thing cannot exist.

\subsection{Summary}

We summarize the relevant lessons learned so far:
\begin{enumerate}
\item[1.] The Weyl character formula for Haar expectations of ratios of characteristic polynomials (c.f.\ Section \ref{sect:WeylCF}) is uniformly valid for \underline{all} $N$ in the supersymmetric case of $n_0 = n_1$ (and $m_0 = m_1$ for type $A$). For that observable, there are no visible phenomena indicating any breakdown for small $N$.
\item[2.] The bosonic color-flavor transformation of type $BD$ in its standard form (\ref{eq:type-A}) does break down for small values of $N$. (Of course, the situation for type $A$ and type $C$ is similar.) Even so, a CF transformation with a modified choice of integration domain and adapted invariant integral may still exist (see Ref.\ \cite{FK} to learn how this goes for type $A$). We have demonstrated that possibility explicitly for the simple case of type $BD$ and $n_1 = 0$, $n_0 = 1$, $N = 2$. (The case of $N = 1$ turned out to be beyond repair.)
\end{enumerate}

Now, the practitioner keen to apply our method will ask about the range of validity of the color-flavor transformation for the super-case ($n_0 = n_1$ for type $C$, $BD$; and $m_0 = n_0 = m_1 = n_1$ for type $A$). That is a major open question. In view of observation 1.\ above, an optimist might have hoped that the CF transformation could be valid for small values of $N$, possibly down to $N \geq 1$; and if it does fail for small $N$ in its standard form, that it could be repaired along the lines of observation 2.\ above. Unfortunately, the matter is complicated, and this author does not know the general answer. In the next section, to develop some intuition, we will make a detailed study of a special case.

\section{Case study: two-point function (type $A$)}\label{sect:TPfctn}
\setcounter{equation}{0}

Let us recall what is at stake. The Wegner-Efetov formalism {\cite{Efetov-Book, Wegner-Book}} employs variables of commuting and anti-commuting type. By the introduction of collective fields, the presence of the former gives rise to a noncompact sector with hyperbolic symmetry \cite{Wegner79}. In the specific case of the color-flavor transformation, one has to deal with the noncompact factor $M_0$ in the domain of the matrix variables $Z, \widetilde{Z}$. As we have indicated, the noncompactness of $M_0$ poses a question, and a non-trivial one at that, as to whether the transformation exists as a rigorous mathematical tool when the number $N$ of colors is small. Here, we shall try to get some insight by inspecting a case of considerable importance for practical applications: $U = \mathrm{U}(N)$ with
\begin{equation}\label{eq:paraset}
    m_0 = n_0 = m_1 = n_1 = 1 .
\end{equation}
That parameter set is used in the supermatrix formalism to compute two-point functions (of the density or density of states). Its representation-theoretic aspects are ruled by a Howe pair $(U, \mathfrak{g})$ of type $A$ with complex Lie superalgebra
\begin{equation}
    \mathfrak{g} = \mathfrak{gl}(2|2) ,
\end{equation}
and the matrices $Z$ and $\widetilde{Z}$ for the integral on the right-hand side of the CF transformation (\ref{eq:type-A}) are $2 \times 2$ supermatrices:
\begin{equation}
    Z = \begin{pmatrix} Z^{00} &Z^{01} \cr Z^{10} &Z^{11} \end{pmatrix} , \quad \widetilde{Z} = \begin{pmatrix} \widetilde{Z}_{00} &\widetilde{Z}_{01} \cr \widetilde{Z}_{10} &\widetilde{Z}_{11} \end{pmatrix} .
\end{equation}
The diagonal matrix entries are complex variables subject to the (anti-)Hermiticity relations $\widetilde{Z}_{00} = + \overline{Z^{00}}$ and $\widetilde{Z}_{11} = - \overline{Z^{11}}$; the off-diagonal matrix entries $Z^{01}$, $Z^{10}$, $\widetilde{Z}_{01}\,$, $\widetilde{Z}_{10}\,$, are Grassmann variables. The domain of integration for $Z$, $\widetilde{Z}$ is a direct product $M = M_0 \times M_1$ of a Poincar\'e disk $M_0$ (parameterized by $|Z^{00}| < 1$) and a Riemann sphere $M_1$ (with complex stereographic coordinate $Z^{11}$ ranging through $\mathbb{C}$). The complex Lie supergroup $G_\mathbb{C} = \mathrm{GL}(2|2)$ acts on $Z$ and $\widetilde{Z}$ by rational transformations:
\begin{eqnarray}
    &&T \cdot Z = \big( T_{\ {\rm a}}^{\rm a} Z + T^{\rm ar} \big) \big( T_{\rm ra} Z + T_{\rm r}^{\ {\rm r}})^{-1} , \cr &&T \cdot \widetilde{Z} = \big( T_{\rm r}^{\ {\rm a}} \widetilde{Z} +  T_{\rm r a} \big) \big( T^{\rm ar} \widetilde{Z} + T_{\ {\rm a}}^{\rm a} )^{-1} ,
\end{eqnarray}
where
\begin{equation}
    T = \begin{pmatrix} T_{\ {\rm a}}^{\rm a} &T^{\rm ar} \cr T_{\rm ra} &T_{\rm r}^{\ {\rm r}} \end{pmatrix}
\end{equation}
is the decomposition of $T \in G_\mathbb{C}$ into blocks for the advanced (a) and retarded (r) sectors. (The varying index positions remind us that vectors are exchanged for co-vectors when we switch between the two sectors.) The elements of a real subgroup $G = \mathrm{U}(1,1|2) \subset G_\mathbb{C}$ preserve the (anti-)Hermiticity relations between $Z$ and $\widetilde{Z}$ and thus act on the (supermanifold over the) domain $M_0 \times M_1\,$. The metric tensor
\begin{equation}
    \mathrm{STr}\, (1-\tilde{Z} Z)^{-1} d\tilde{Z}\, (1-Z\tilde{Z})^{-1} dZ
\end{equation}
is $G$-invariant. According to Eq.\ (\ref{eq:Dm-typeA}) for the parameter set (\ref{eq:paraset}), the $G$-invariant Berezin integral form derived from it is ``flat'':
\begin{equation}
    D\mu(Z,\widetilde{Z}) = D(Z,\widetilde{Z}) = d^2 Z^{11} d^2 Z^{00} \; \partial_{Z,\tilde{Z}} \,,
\end{equation}
where
\begin{equation}
    \partial_{Z,\tilde{Z}} \equiv \frac{\partial^2}{\partial {Z}^{01} \partial\widetilde{Z}_{10}} \, \frac{\partial^2}{\partial\widetilde{Z}_{01} \partial Z^{10}} \,,
\end{equation}
and the area elements are normalized so that
\begin{equation}
    \int_{|Z^{00}| < 1} \!\!\!\!\!\! d^2 Z^{00} = \pi = \int_\mathbb{C} d^2 Z^{11} \, \left( 1 + |Z^{11}|^2 \right)^{-2} .
\end{equation}
Introducing single-particle operators for the advanced ($a$) and retarded ($c$) sectors, we define the overlap kernel for spin-coherent states as
\begin{equation}\label{eq:Okern}
    K_N(\widetilde{Z},Z^\prime) = \langle 0 \vert \exp \big( c^{\nu j} (-1)^{|\nu|} \widetilde{Z}_{\nu\mu} a_{\ j}^{\, \mu} \big) \exp \big( \bar{a}_{\ \mu}^{\, i} {Z^\prime}^{\mu\nu} \bar{c}_{i \nu} \big) \vert 0 \rangle \,.
\end{equation}
By the general commutation relations (\ref{eq:CAR}, \ref{eq:CCR}) this has the expression
\begin{equation}
    K_N(\widetilde{Z},Z^\prime) = \mathrm{SDet}^{-N}(1-\widetilde{Z} Z^\prime) .
\end{equation}
In order for the CF transformation (\ref{eq:type-A}) to hold true, we need the kernel to be reproducing:
\begin{equation}\label{eq:repro}
    \int_M D\mu_N(Z,\widetilde{Z}) \, K_N(\widetilde{Z}^\prime,Z)\, K_N (\widetilde{Z} , Z^{\prime\prime}) \stackrel{?}{=} K_N(\widetilde{Z}^\prime, Z^{\prime\prime}) ,
\end{equation}
with integration measure
\begin{equation}
    D\mu_N(Z,\widetilde{Z}) \equiv c_A^{(N)} D(Z,\widetilde{Z}) \circ \mathrm{SDet}^N (1-\widetilde{Z} Z)
\end{equation}
for the integration domain $M = M_0 \times M_1$ given by $|Z^{00}| < 1$ and $|Z^{11}| < \infty$, and with a normalization constant $c_A^{(N)}$ that is neither zero nor infinite.

\subsection{First Check}\label{sect:1stfail}

First of all, we work out the normalization integral that results from setting $\widetilde{Z}^\prime = Z^{\prime\prime} = 0$ in Eq.\ (\ref{eq:repro}):
\begin{equation}\label{eq:check0}
    \int_M D\mu_N(Z,\widetilde{Z}) \stackrel{?}{=} K_N(0,0) = 1 .
\end{equation}
To compute the integral on the left-hand side, we introduce the following notation:
\begin{equation}\label{eq:expS024}
    \mathrm{SDet}\,(1 - \tilde{Z} Z) = S_0 + S_2 + S_4 \,,
\end{equation}
where
\begin{eqnarray}
    &&S_0 = \frac{1- \widetilde{Z}_{00} Z^{00}}{1 - \widetilde{Z}_{11} Z^{11}} \,, \cr
    &&S_2 = - \frac{ \widetilde{Z}_{01} Z^{10} + Z^{01} \widetilde{Z}_{10} + \widetilde{Z}_{01} Z^{11} \widetilde{Z}_{10} Z^{00} + Z^{01} \widetilde{Z}_{11} Z^{10} \widetilde{Z}_{00}}{(1 - \widetilde{Z}_{11} Z^{11})^2} \,,\cr
    &&S_4 = \widetilde{Z}_{01} Z^{10} Z^{01} \widetilde{Z}_{10} \frac{1 + \widetilde{Z}_{11} Z^{11}}{(1 - \widetilde{Z}_{11} Z^{11})^3} \,.
\end{eqnarray}
Then by binomial expansion we have
\begin{equation}
    \mathrm{SDet}^N(1 - \widetilde{Z} Z) = S_0^N + N S_0^{N-1} (S_2 + S_4) + {\textstyle{\frac{1}{2}}} N(N-1)\, S_0^{N-2} S_2^2 \,.
\end{equation}
It follows that
\begin{eqnarray}
    \partial_{Z,\widetilde{Z}} \, \mathrm{SDet}^N(1 - \widetilde{Z} Z) &=&
    N S_0^{N-1} \frac{1 + \widetilde{Z}_{11} Z^{11}}{(1 - \widetilde{Z}_{11} Z^{11})^3} \\ &+& N(N-1) S_0^{N-2} \, \frac{1 - \widetilde{Z}_{00} Z^{00} \widetilde{Z}_{11} Z^{11}}{(1 - \widetilde{Z}_{11} Z^{11})^4} \,. \nonumber
\end{eqnarray}
Now, introducing the ``radial'' variables $t_0 = \vert Z^{00} \vert^2$ and $t_1 = \vert Z^{11} \vert^2$ we obtain
\begin{eqnarray}\label{eq:norm-17}
    \int_M D\mu_N(Z,\widetilde{Z}) &=& c_A^{(N)} \pi^2 \int\limits_{0}^\infty dt_1 \int\limits_0^1 dt_0 \left( N \left( \frac{1 - t_0}{1 + t_1} \right)^{N-1} \frac{1 - t_1}{(1 + t_1)^3} \right. \cr &+& \left. N(N-1) \left( \frac{1 - t_0}{1 + t_1} \right)^{N-2} \frac{1 + t_0 t_1}{(1 + t_1)^4} \right).
\end{eqnarray}
The right-hand side can be expressed as a sum of products of the elementary integrals $\int_0^1 (1-t_0)^p dt_0 = (p+1)^{-1}$ and $\int_0^\infty (1+t_1)^{-q} dt_1 = (q-1)^{-1}$ ($q > 1$). In this way we easily find
\begin{equation}
    \int_M D\mu_N(Z,\widetilde{Z}) = c_A^{(N)} \pi^2 .
\end{equation}
Thus Eq.\ (\ref{eq:check0}) holds with an $N$-independent normalization constant, $c_A^{(N)} = \pi^{-2}$.

Superficially seen, everything looks fine and there appears to be no indication of any problems for small $N$. That, however, is deceiving. In fact, for $N = 1$ the second line in (\ref{eq:norm-17}) vanishes and we get
\begin{equation}\label{eq:norm0}
    \int_M D\mu_{1}(Z,\widetilde{Z}) = c_A^{(N)} \pi^2 \int\limits_0^1 dt_0 \int\limits_{0}^\infty dt_1 \frac{1 - t_1}{(1 + t_1)^3} = 0 \not= 1 .
\end{equation}
(The calculation above used $(N-1)/(N-1) = 1$ as an intermediate step, illegal for $N = 1$.) Thus the normalization integral for $N = 1$ is actually zero and cannot be set to unity with a finite normalization constant $c_{N=1}^{(A)}$.

\subsection{Second check}\label{sect:2ndfail}

For a second check, we linearize the reproducing property (\ref{eq:repro}) with respect to both $\widetilde{Z}^\prime$ and $Z^{\prime\prime}$. The resulting relation is
\begin{eqnarray}\label{eq:check1}
    &&I_{\nu_2 \mu_2}^{\mu_1 \nu_1} (N) \equiv \int_M D\mu_N(Z,\widetilde{Z})
    \, Z^{\mu_1 \nu_1} \widetilde{Z}_{\nu_2 \mu_2} \stackrel{?}{=} N^{-1} (-1)^{|\nu_1|} \delta_{\nu_2}^{\nu_1} \delta_{\mu_2}^{\mu_1} \,.
\end{eqnarray}
Here let us recall our notation for fermion parity: $|\nu| = 0$ (even or bosonic index) and $|\nu| = 1$ (odd or fermionic index). All of the cases in Eq.\ (\ref{eq:check1}) work out perfectly with normalization constant $c_N^{(A)} = \pi^{-2}$, as long as $N \geq 2$. For example,
\begin{eqnarray}
    I_{11}^{11} (N) &=& - \int\limits_{0}^\infty dt_1 \int\limits_0^1 dt_0 \left( N \left( \frac{1 - t_0}{1 + t_1} \right)^{N-1} \frac{t_1 - t_1^2}{(1 + t_1)^3} \right. \\ &+& \left. N(N-1) \left( \frac{1 - t_0}{1 + t_1} \right)^{N-2} \frac{t_1 + t_0 t_1^2}{(1 + t_1)^4} \right) = - \frac{1}{N} \,, \nonumber
\end{eqnarray}
as is easily verified. Also,
\begin{eqnarray}
    I_{00}^{00} (N) &=& \int\limits_{0}^\infty dt_1 \int\limits_0^1 dt_0 \left( N \left( \frac{1 - t_0}{1 + t_1} \right)^{N-1} \frac{t_0(1 - t_1)}{(1 + t_1)^3} \right. \\ &+& \left. N(N-1) \left( \frac{1 - t_0}{1 + t_1} \right)^{N-2} \frac{t_0 + t_0^2 t_1}{(1 + t_1)^4} \right) = + \frac{1}{N} \,. \nonumber
\end{eqnarray}
To compute $I_{01}^{10} (N)$ note that
\begin{equation}
    \partial_{Z,\widetilde{Z}} \, Z^{01} \, \mathrm{SDet}^N(1 - \widetilde{Z} Z) \, \widetilde{Z}_{10} = N S_0^{N-1} \big( 1 - \widetilde{Z}_{11} Z^{11} \big)^{-2} .
\end{equation}
Hence
\begin{equation}
    I_{01}^{10}(N) = N \int\limits_{0}^\infty dt_1 \int\limits_0^1 dt_0
    \frac{(1-t_0)^{N-1}}{(1+t_1)^{N+1}} = \frac{1}{N} \,,
\end{equation}
as required. Similarly, $I_{10}^{01}(N) = -1/N$, as required.

All other integrals $I_{\nu_2 \mu_2}^{\mu_1 \nu_1} (N)$ vanish on symmetry grounds. Everything still appears to be in good shape (uniformly in $N$). However, for $N = 1$ there is again a problem: the integral
\begin{equation}\label{eq:I11-bulk}
    I_{11}^{11} (N=1) = \int\limits_0^1 dt_0 \int\limits_{0}^\infty dt_1  \frac{t_1^2 - t_1}{(1 + t_1)^3} = \infty
\end{equation}
diverges instead of being $-1$. Similar problems recur in higher order (of $\widetilde{Z}^\prime Z^{\prime\prime}$). Thus the type-$A$ color-flavor transformation with parameter set (\ref{eq:paraset}) is sick for $N = 1$. Is there a remedy for that problem? The answer is yes, as we shall see in the next subsection.

\subsection{Correction term for $N=1$}\label{sect:fix}

In this subsection, let $N = 1$ throughout. Our inspiration here comes from Section \ref{sect:3.8} where we saw that, while the integral on the right-hand side of the bosonic CF transformation of type $BD$ ceases to exist for low values of $N$, one can correct the  transformation (in that case for $N = 2$) by moving the weight of the integral from the ``bulk'' of the noncompact Poincar\'e disk $|z| < 1$ to its ``surface'', $|z| = 1$. In the present case, the corrected expression will turn out to be a linear combination of bulk and surface contributions.

Parameterizing $|Z^{00}| = 1$ by $Z^{00} = \mathrm{e}^{\mathrm{i} \theta}$ (and $\widetilde{Z}_{00} = \mathrm{e}^{- \mathrm{i}\theta}$), consider the following Berezin integral localized on the surface $M_1 \times \partial M_0\,$:
\begin{equation}\label{eq:Isurf}
    I_{\rm surf} [F] = \pi^{-2} \int_{\mathbb{C}} d^2 Z^{11} \int_0^{2\pi} \!\!\!\! d\theta \; \partial_{Z,\widetilde{Z}} \left( \Omega \; \mathrm{SDet}(1-\widetilde{Z} Z) F \right) \Big\vert_{Z^{00} = \mathrm{e}^{\mathrm{i}\theta}} \,,
\end{equation}
where
\begin{eqnarray}\label{eq:Omega}
    \Omega &=& - \frac{ \widetilde{Z}_{01} Z^{10} + Z^{01} \widetilde{Z}_{10} + \widetilde{Z}_{01} Z^{11} \widetilde{Z}_{10} Z^{00} + Z^{01} \widetilde{Z}_{11} Z^{10} \widetilde{Z}_{00} }{2(1 - \widetilde{Z}_{11} Z^{11})} \cr &+& \widetilde{Z}_{01} Z^{10} Z^{01} \widetilde{Z}_{10} \left( \frac{1}{2(1 - \widetilde{Z}_{11} Z^{11})^2} + \frac{1}{4(1 - \widetilde{Z}_{11} Z^{11})} \frac{\partial}{\partial |Z^{00}|} \right) .
\end{eqnarray}
Notice the peculiar feature that $\Omega$ is not a function but a differential operator: the derivative $\partial / \partial |Z^{00}|$ normal to the surface is applied to the integrand as a function on the domain $|Z^{00}| \leq 1$ before the restriction to the boundary ($|Z^{00}| =  1$) is made. (Of course, the partial derivative is defined w.r.t.\ the local coordinate system given by the $Z^{\mu\nu}$ and $\widetilde{Z}_{ \nu\mu}$.)

Our claim now is that the reproducing property (\ref{eq:repro}) with parameter set (\ref{eq:paraset}) still holds for $N = 1$ if the bulk integral over $M$,
\begin{equation}
    I_{\rm bulk}[F] \equiv \int_M D\mu_{1}(Z,\widetilde{Z}) \, F ,
\end{equation}
with Berezin integral form $D \mu_1(Z,\widetilde{Z}) = c_A^{(1)} D(Z,\widetilde{Z}) \, \mathrm{SDet}(1-\widetilde{Z} Z)$ and normalization constant $c_A^{(1)} = \pi^{-2}$, is augmented by the surface integral (\ref{eq:Isurf}):
\begin{equation}
    I[F] = (I_{\rm bulk} + I_{\rm surf})[F] .
\end{equation}
We begin by checking the cases where Sections \ref{sect:1stfail}, \ref{sect:2ndfail} detected a failure for $N = 1$. Since we already know that $I_{\rm bulk}[1] = 0$, the normalization integral shortens to
\begin{equation}
    I[F \equiv 1] = I_{\rm surf}[1] = \pi^{-2} \int_{\mathbb{C}} d^2 Z^{11} \int_0^{2\pi} \!\!\!\! d\theta \; \partial_{Z,\widetilde{Z}} \, \Omega \, (S_0 + S_2) \Big\vert_{Z^{00} = \mathrm{e}^{\mathrm{i} \theta}} \,.
\end{equation}
Here the term $S_4$ in the expansion (\ref{eq:expS024}) of $\mathrm{SDet}(1-\widetilde{Z} Z)$ has been dropped because it is of maximal degree in the Grassmann variables and $\Omega$ contains no term of degree zero. Now there are two contributions to the normalization integral, one from the summand $S_0$ and one from $S_2$. The contribution from $S_2$ is
\begin{eqnarray*}
    &&\int_{\mathbb{C}} d^2 Z^{11} \int_0^{2\pi} \!\!\!\! d\theta \; \partial_{Z,\widetilde{Z}} \; \frac{ (\widetilde{Z}_{01} Z^{10} + Z^{01} \widetilde{Z}_{10} + \widetilde{Z}_{01} Z^{11} \widetilde{Z}_{10} \mathrm{e}^{\mathrm{i}\theta} + Z^{01} \widetilde{Z}_{11} Z^{10} \mathrm{e}^{- \mathrm{i}\theta})^2 }{2 \pi^2 (1 - \widetilde{Z}_{11} Z^{11})^3} \cr &&= \frac{1}{\pi^2} \int d\theta \int \frac{d^2 Z^{11}}{(1-\widetilde{Z}_{11} Z^{11})^2} =
    \frac{2}{\pi} \int_0^\infty \frac{\pi dt_1}{(1+t_1)^2} = 2 .
\end{eqnarray*}
In the summand containing the factor $S_0\,$, which vanishes on the boundary $|Z^{00}| = 1$, only the derivative term in $\Omega$ gives a nonzero result:
\begin{equation*}
    \int_{\mathbb{C}} d^2 Z^{11} \int_0^{2\pi} \!\!\!\! d\theta \; \partial_{Z,\widetilde{Z}} \; \frac{\widetilde{Z}_{01} Z^{10} Z^{01} \widetilde{Z}_{10}} {4\pi^2 (1 - \widetilde{Z}_{11} Z^{11})} \frac{\partial}{\partial |Z^{00}|}\, S_0 = -\int_0^\infty \frac{dt_1}{(1+t_1)^2} = -1.
\end{equation*}
Altogether, we find the result needed for the reproducing property (\ref{eq:repro}) at $\widetilde{Z}^\prime = Z^{\prime\prime} = 0$:
\begin{equation}\label{eq:I[1]}
    I[1] = I_{\rm surf}[1] = 2 - 1 = 1 .
\end{equation}

Second, let us check the integral $I[\widetilde{Z}_{11} Z^{11}]$, which ought to be $-1$ according to Eq.\ (\ref{eq:check1}) but came out as $I_{\rm bulk}[\widetilde{Z}_{11} Z^{11}] = \infty$ (Section \ref{sect:2ndfail}). The integrand for $I_{\rm surf}[\widetilde{Z}_{11} Z^{11}]$ is the integrand for $I[1]$ multiplied by $\widetilde{Z}_{11} Z^{11} = - t_1$:
\begin{equation}
    I_{\rm surf}[\widetilde{Z}_{11} Z^{11}] = - \int_0^\infty \frac{t_1 dt_1}{(1+t_1)^2} = \infty .
\end{equation}
Adding this to the bulk contribution (\ref{eq:I11-bulk}) we obtain
\begin{equation}\label{eq:4.35}
    (I_{\rm bulk} + I_{\rm surf})[\widetilde{Z}_{11} Z^{11}] = \int_0^\infty dt_1 \left( \frac{t_1^2 - t_1}{(1+t_1)^3} - \frac{t_1}{(1+t_1)^2} \right) = -1 .
\end{equation}
Thus although $I_{\rm bulk}[\bullet]$ and $I_{\rm surf}[\bullet]$ do not exist individually on $\bullet = \widetilde{Z}_{11} Z^{11}$, their sum does exist and has the required value, $-1$.

In summary, our modified integral expression $I = I_{\rm bulk} + I_{\rm surf}$ has passed some obvious tests. In the next subsection, we shall prove that it indeed corrects the color-flavor transformation for the case at hand. Still, the puzzled reader might ask: where did the formula (\ref{eq:Omega}) for $\Omega$ come from? Let us therefore indicate briefly how that formula can be constructed. The main idea is that there exist local coordinate systems which make the $|Z^{00}| = 1$ boundary singularity disappear and re-appear at another location, where it can be handled more easily (e.g., the coordinates given by a Cartan-polar decomposition $G = K A K$ are of that kind). Starting from such coordinates, one applies the change-of-variables formula \cite{Rothstein} for integration on noncompact supermanifolds. We will not go through that tedious procedure here but simply verify the outcome (\ref{eq:Omega}) by the means developed in the present paper.

\subsection{$N=1$ type-$A$ CF transformation: proof}\label{sect:proof}

We shall now prove that the type-$A$ color-flavor transformation (\ref{eq:type-A}) with parameter set (\ref{eq:paraset}) extends to the pathological case of $N=1$, by the redefinition of the right-hand side as the boundary-corrected integral $I = I_{\rm bulk} + I_{\rm surf}\,$. We will do so by verifying the reproducing property of the overlap kernel $K_{N=1}\,$. To begin, recall the definition of that kernel:
\begin{equation}
    K_1(\widetilde{Z},Z) = \langle 0 \vert \exp \big( c^{\nu} (-1)^{|\nu|} \widetilde{Z}_{\nu\mu} a^{\mu} \big) \exp \big( \bar{a}_{\mu} Z^{\mu\nu} \bar{c}_{\nu} \big) \vert 0 \rangle \,,
\end{equation}
which is Eq.\ (\ref{eq:Okern}), simplified by the specialization $N = 1$. Next, we expand the overlap kernel by inserting a complete set of orthonormal states, $P = \sum \vert v \rangle \langle v |\,$, for the projector onto the subspace of color-neutral states in Fock space. This results in a sum,
\begin{equation}
    K_1(\widetilde{Z},Z) = \sum_v e_v(\widetilde{Z}) f^v(Z) ,
\end{equation}
with factors given by
\begin{equation}
    e_v(\widetilde{Z}) = \langle 0 \vert \exp \big( c^{\nu} (-1)^{|\nu|} \widetilde{Z}_{\nu\mu} a^{\mu} \big) \vert v \rangle , \quad f^v(Z) = \langle v \vert \exp \big( \bar{a}_{\mu} Z^{\mu\nu} \bar{c}_{\nu} \big) \vert 0 \rangle \,.
\end{equation}
Mathematically speaking, these factors are (anti-)holomorphic sections of a $\mathbb{C}$-line bundle and its dual. Denoting the number of particle pairs in the many-body state $\vert v \rangle$ by $k$ and making some natural choices, we arrive at the list of sections displayed in Table \ref{table:sections}.
\begin{table}
\begin{center}
\begin{tabular}{l|l|l}
$f^v(Z)$   &$e_v(\widetilde{Z})$ &index range\\
\hline
${(Z^{00})^k}^{\vphantom{\dagger}}$  &$(\widetilde{Z}_{00})^k$ &$k \geq 0$\\
$-\sqrt{k}\, (Z^{00})^{k-1} Z^{01}$ &$\sqrt{k}\,\widetilde{Z}_{10}(\widetilde{Z}_{00})^{k-1}$ &$k\geq 1$\\
$\sqrt{k}\, (Z^{00})^{k-1} Z^{10}$ &$\sqrt{k}\,\widetilde{Z}_{01}(\widetilde{Z}_{00})^{k-1}$ &$k \geq 1$\\
$(Z^{00})^{l-1} (Z^{00} Z^{11} - l Z^{01} Z^{10})$ &$(l \widetilde{Z}_{01} \widetilde{Z}_{10} - \widetilde{Z}_{11} \widetilde{Z}_{00}) (\widetilde{Z}_{00})^{l-1}$ &$l \equiv k-1 \geq 0$
\end{tabular}
\end{center}
\caption{List of (anti-)holomorphic line-bundle sections and their duals ($N=1$).} \label{table:sections}
\end{table}

Clearly, the first line of the table stems from $k$ boson-boson pairs (each consisting of one retarded and one advanced particle), the second and third lines from one boson-fermion (or fermion-boson) and $k-1$ boson-boson pairs, and the fourth line from one fermion-fermion and $k-1$ boson-boson pairs (which is degenerate, due to $N=1$, with $k-2$ boson-boson pairs together with one boson-fermion and one fermion-boson pair). The fourth-line entries for $k=1$ are to be understood as single fermion-fermion pairs.

The reproducing property of the overlap kernel $K_1$ (hence the CF transformation for $N=1$) is proved by checking the orthogonality relations
\begin{equation}
    I \big[  f^v(Z) \, e_w(\widetilde{Z}) \big] = \delta_{w}^v
\end{equation}
for the modified integral $I = I_{\rm bulk} + I_{\rm surf}\,$. Now our integral $I$ has an obvious $K_L \times K_R$ symmetry $Z \mapsto k_L Z k_R^{-1}$, $\widetilde{Z} \mapsto k_R \widetilde{Z} k_L^{-1}$ for $K_L = K_R = \mathrm{U}(1) \times \mathrm{U}(1) \subset \mathrm{U}(1|1)$. Different terms in the same column of our table belong to different representations of $K_L \times K_R$.
Therefore, all off-diagonal terms ($v \not= w$) vanish trivially, and it suffices to check the diagonal: \begin{equation}
     I \big[  f^v(Z) \, e_v(\widetilde{Z}) \big] \stackrel{?}{=} 1 .
\end{equation}
In the following, ``degree'' always means the degree in the Grassmann algebra with generators $Z^{01}$, $Z^{10}$, $\widetilde{Z}_{01}\,$, and $\widetilde{Z}_{10}\,$.

Beginning with the terms in the first line of Table \ref{table:sections}, we recall from Section \ref{sect:1stfail} that $I_{\rm bulk}[1] = 0$. By the same token, $I_{\rm bulk} \big[ F_1 \equiv (Z^{00} \widetilde{Z}_{00} )^k \big] = 0$. Hence,
\begin{eqnarray}
    &&I\big[ (Z^{00} \widetilde{Z}_{00} )^k \big] = I_{\rm surf} \big[ F_1 \big] \cr
    &&= \pi^{-2} \int_{\mathbb{C}} d^2 Z^{11} \int_0^{2\pi} \!\!\!\! d\theta \; \partial_{Z,\widetilde{Z}} \left( \Omega \; \mathrm{SDet}(1-\widetilde{Z} Z) \vert Z^{00} \vert^k \right) \Big\vert_{Z^{00} = \mathrm{e}^{\mathrm{i}\theta}} \,.
\end{eqnarray}
Now with reference to Eq.\ (\ref{eq:expS024}), we replace $\mathrm{SDet}(1-\widetilde{Z} Z)$
by $S_0 + S_2\,$; the degree-4 summand $S_4$ can be omitted because the surface factor $\Omega$ raises the degree by no less than 2 (and the Fermi integral $\partial_{Z,\widetilde{Z}}$
picks out degree 4). Moreover, since the derivative in $\Omega$ comes with degree 4, it has to be paired with $S_0\,$, whence it is consumed by the factor $1-|Z^{00}|^2$, vanishing for $Z^{00} = \mathrm{e}^{\mathrm{i} \theta}$. Therefore, we may replace the bulk integrand $F_1 = |Z^{00}|^k$ by its surface value $|\mathrm{e}^{\mathrm{i} \theta}|^k = 1$. According to Eq.\ (\ref{eq:I[1]}) this yields
\begin{equation}
    I\big[ (Z^{00} \widetilde{Z}_{00} )^k \big] = I_{\rm surf}[1] = 1 ,
\end{equation}
as required.

We turn to the terms in the second line of Table \ref{table:sections}, i.e., we compute $I[F_2]$ for
\begin{equation}\label{eq:2ndline}
    F_2 = f^v(Z) e_v(\widetilde{Z}) = - k Z^{01} \widetilde{Z}_{01} |Z^{00}|^{2k-2} .
\end{equation}
Here the surface contribution vanishes identically. Indeed, since $F_2$ in (\ref{eq:2ndline}) has degree 2, we must select from $\Omega$ the term of minimal degree (which is 2) and from $\mathrm{SDet}(1-\widetilde{Z} Z)$ the degree-$0$ term $S_0$ (so as to arrive at degree 4), but $S_0$ vanishes on the surface. Hence
\begin{equation}
    I[F_2] = I_{\rm bulk}[F_2] = \pi^{-2} \int_M D(Z,\widetilde{Z}) \, \mathrm{SDet}(1 - \widetilde{Z} Z) \left( - k Z^{01} \widetilde{Z}_{01} |Z^{00}|^{2k-2} \right) .
\end{equation}
Now, again for reasons of degree, we can replace $\mathrm{SDet}(1 - \widetilde{Z} Z)$ by its constituent $S_2\,$. Carrying out the Fermi integral $\partial_{Z,\widetilde{Z}}$ and changing integration variables to $t_0 = |Z^{00}|^2$ and $t_1 = |Z^{11}|^2$, we then obtain the required result:
\begin{equation}
    I\big[  - k Z^{01} \widetilde{Z}_{01} |Z^{00}|^{2k-2} \big] = k \int_0^\infty dt_1 \int_0^1 dt_0 \, \frac{t_0^{k-1}}{(1+t_1)^2} = 1 .
\end{equation}
The case of the terms in the third line of Table \ref{table:sections} is no different; we therefore skip it.

The terms in the fourth line,
\begin{equation}
    F_4 \equiv - (Z^{00} Z^{11} - l Z^{01} Z^{10}) (\widetilde{Z}_{11} \widetilde{Z}_{00} - l \widetilde{Z}_{01} \widetilde{Z}_{10}) |Z^{00}|^{2l-2} ,
\end{equation}
require more effort. This is because their bulk and surface integrals are divergent individually and, as we saw earlier for the special case of $F_4 = -Z^{11} \widetilde{Z}_{11}$ ($l=0$), they must be combined to arrive at a finite result. [Since the case of $l=0$ has already been dealt with in Section \ref{sect:fix}, Eq.\ (\ref{eq:4.35}), we may assume $l \geq 1$.] We first work out the bulk integral. Here all summands in $\mathrm{SDet}\,(1-\widetilde{Z} Z) = S_0 + S_2 + S_4$ contribute. The contribution from the summand $S_0$ is
\begin{equation}
    I_{\rm bulk}^{(S_0)}[F_4] = - \frac{l^2}{\pi^2} \int D(Z,\widetilde{Z}) \, S_0 Z^{01} Z^{10} \widetilde{Z}_{01} \widetilde{Z}_{10} |Z^{00}|^{2l-2} .
\end{equation}
This comes out as
\begin{equation}
    I_{\rm bulk}^{(S_0)}[F_4] = l^2 \int dt_1 \int dt_0 \, \frac{1-t_0}{1+t_1} \, t_0^{l-1}
    = \frac{l}{l+1} \int \frac{dt_1}{1+t_1} \,.
\end{equation}
Next, the summand $S_2$ singles out the cross terms, namely $Z^{11} Z^{00} \widetilde{Z}_{01} \widetilde{Z}_{10}$ and $Z^{01} Z^{10} \widetilde{Z}_{00} \widetilde{Z}_{11}$, each multiplied by $l |Z^{00}|^{2l-2}$, in $F_4\,$. For these we obtain
\begin{equation}
    I_{\rm bulk}^{(S_2)}[F_4] = ... = - \frac{2l}{l+1} \int \frac{t_1 dt_1}{(1+t_1)^2} \,.
\end{equation}
Finally, the summand $S_4$ pairs with $|Z^{11}|^2 |Z^{00}|^{2l}$ in $F_4$ to give the contribution
\begin{equation}
    I_{\rm bulk}^{(S_4)}[F_4] = ... = \frac{1}{l+1} \int dt_1 \, \frac{t_1 -t_1^2}{(1+t_1)^3} \,.
\end{equation}
By adding all three contributions, we get the following result for the bulk integral:
\begin{equation}
    I_{\rm bulk}[F_4] = \int dt_1 \, \frac{l + t_1 - (l+1)\, t_1^2}{(l+1) (1+t_1)^3} \,.
\end{equation}

We turn to the surface integral. Here we obtain
\begin{equation}
     I_{\rm surf}^{(S_0)}[F_4] = \int d^2 Z^{11} d\theta \, \partial_{Z,\widetilde{Z}}
     \frac{\widetilde{Z}_{01} Z^{10} Z^{01} \widetilde{Z}_{10}}{4\pi^2 (1-\widetilde{Z}_{11} Z^{11})} F_4 \frac{\partial}{\partial |Z^{00}|} S_0 = \int \frac{- t_1 dt_1}{(1+t_1)^2} \,,
\end{equation}
and
\begin{equation}
     I_{\rm surf}^{(S_2)}[F_4] = ... = \int \frac{2 t_1 dt_1}{(1+t_1)^2} \,,
\end{equation}
where the last computation is essentially the same as that for $F_4 = - Z^{11} \widetilde{Z}_{11}$ in Sect.\ \ref{sect:fix}. The summand $S_4$ cannot contribute, again for the degree reason. Altogether, we thus arrive at
\begin{equation}
    I[F_4] = (I_{\rm bulk} + I_{\rm surf}) [F_4] = \int dt_1 \, \frac{l + (l+2) t_1}{(l+1) (1+t_1)^3} = 1 ,
\end{equation}
which is the result that had to be shown.

Here ends our tour-de-force proof. It is clear that the present kind of approach by explicit calculation will hardly be viable for higher values of the parameters $m_0$, $m_1$, $n_0$, $n_1\,$. Nonetheless, lacking a more conceptual approach we have presented it here, if only to issue a warning of what kind of phenomena may occur when one tries to apply the color-flavor transformation outside its standard range of validity.

\subsection{New alternative method}

In the previous section, we saw that the type-$A$ color-flavor transformation for the case of a single color ($N = 1$) and one flavor replica ($m_0 = m_1 = n_0 = n_1 = 1$) is a delicate matter; from a practical viewpoint it is rather unwieldy, as the standard bulk integral must be augmented by a non-standard surface integral. That state of affairs motivates us to seek an alternative approach. Actually, such an approach has been available for a while: it is suggested by the collection of formulas derived in Section 3 of Ref.\ \cite{BWZ}. Although that approach applies in great generality, we describe it here only for the case at hand (i.e.\ for type $A$, with $m_0 = m_1 = n_0 = n_1 = 1$ and $N = 1$).

\subsubsection{Green's functions via Cayley transform}

Let us recall our goal. Given a product $U = U_d U_f$ of two unitary operators ($U_d$  deterministic and $U_f$ random) acting on a Hilbert space $\mathcal{H}$, we wish to express disorder-averaged products of retarded and advanced Green's functions for $U$ in a good way; i.e., after taking the average over the randomness in $U_f$, we want to be left with a manageable expression, amenable to further processing by analytical means.

We begin with a few elementary formulas. If $g$, $h$ are two operators on $\mathcal{H}$ with the property that each of $1-g$, $1-h$, and $1-gh$, is invertible, we can write
\begin{eqnarray}
    1-gh &=& \frac{1}{2} (1+g)(1-h) + \frac{1}{2} (1-g)(1+h) \cr &=& \frac{1}{2} (1-g) \left( \frac{1+g}{1-g} + \frac{1+h}{1-h} \right) (1-h) ,
\end{eqnarray}
and by taking the inverse on both sides we obtain
\begin{equation}\label{eq:Cay-Id}
    (1-gh)^{-1} = (1-h)^{-1} \left( {\textstyle{\frac{1}{2}}} (A_g + A_h) \right)^{-1} (1-g)^{-1} ,
\end{equation}
where $g \mapsto A_g$ stands for the Cayley transform
\begin{equation}
    A_g = \frac{1+g}{1-g} \,.
\end{equation}
For a contraction $g$ (i.e., $\parallel\! g \!\parallel < 1$), the latter has the important property that
\begin{equation}\label{eq:C-pos}
    \mathrm{Re} \, A_g \equiv \frac{1}{2} \left( A_g + A_g^\dagger \right) > 0 \quad (\parallel \! g \! \parallel < 1) .
\end{equation}
We also note that simple manipulations put Eq.\ (\ref{eq:Cay-Id}) in either one of two adapted forms:
\begin{equation}\label{eq:Cay-Idh}
    (1-gh)^{-1} = (1-h)^{-1} - (1-h)^{-1} \left( {\textstyle{\frac{1}{2}}} (A_g + A_h) \right)^{-1} h (1-h)^{-1}
\end{equation}
and
\begin{equation}\label{eq:Cay-Idg}
    (1-gh)^{-1} = (1-g)^{-1} - g (1-g)^{-1} \left( {\textstyle{\frac{1}{2}}} (A_g + A_h) \right)^{-1} (1-g)^{-1} .
\end{equation}

With these preparations made, we set up the retarded Green's function. For that, we take $g = \alpha U_d$ and $h = \beta U_f$ with two complex numbers $\alpha$ and $\beta$, each of modulus less than unity. We also abbreviate $\alpha\beta \equiv \zeta$. Working in a basis $\vert j \rangle$ of $\mathcal{H}$ in which $U_f$ is diagonal, we write $\mathrm{e}^{ \mathrm{i} \theta_j}$ for the diagonal matrix entries of $U_f$. We then apply Eq.\ (\ref{eq:Cay-Idh}) to express the matrix element of the retarded resolvent between two orthogonal states $|j\rangle$ and $\vert k \rangle$ as
\begin{equation}
    \langle j \vert (1- \zeta U_d U_f)^{-1} \vert k \rangle = \frac{-1}{1 - \beta \mathrm{e}^{\mathrm{i} \theta_j}} \langle j \vert \left( {\textstyle{\frac{1}{2}}} (A_{\alpha U_d} + A_{\beta U_f}) \right)^{-1} \vert k \rangle \frac{\beta \mathrm{e}^{\mathrm{i} \theta_k}}{1 - \beta \mathrm{e}^{\mathrm{i} \theta_k}} \,.
\end{equation}
To express the corresponding matrix element of the advanced resolvent, we change the identifications to $g = \bar\beta U_f^{-1}$ and $h = \bar\alpha U_d^{-1}$, and we then apply Eq.\ (\ref{eq:Cay-Idg}) to obtain
\begin{equation}
    \langle k \vert (1-  \bar\zeta U_f^{-1} U_d^{-1})^{-1} \vert j \rangle = \frac{\bar\beta \mathrm{e}^{-\mathrm{i} \theta_k}}{1 - \bar\beta \mathrm{e}^{-\mathrm{i} \theta_k}} \langle k \vert \left( {\textstyle{\frac{1}{2}}} (A_{\bar\alpha U_d^{-1}} + A_{\bar\beta U_f^{-1}}) \right)^{-1} \vert j \rangle \frac{-1}{1 - \bar\beta \mathrm{e}^{-\mathrm{i} \theta_j}} \,.
\end{equation}
Finally, we multiply the expressions for the retarded and advanced Green's functions to arrive at the formula
\begin{eqnarray}\label{eq:GF-Cayley}
    &&\left\vert \langle j \vert (1- \zeta U_d U_f)^{-1} \vert k \rangle \right\vert^2 =
    \frac{|\beta|^2}{|1 - \beta \mathrm{e}^{\mathrm{i} \theta_j}|^2 \; |1 - \beta \mathrm{e}^{\mathrm{i} \theta_k}|^2} \cr &&\times \langle j \vert \left( {\textstyle{\frac{1}{2}}} (A_{\alpha U_d} + A_{\beta U_f}) \right)^{-1} \vert k \rangle \langle k \vert \left( {\textstyle{\frac{1}{2}}} (A_{\bar\alpha U_d^{-1}} + A_{\bar\beta U_f^{-1}}) \right)^{-1} \vert j \rangle \,.
\end{eqnarray}
Its attractive feature is that the Cayley transform has separated the random factor $U_f$ from the deterministic factor $U_d\,$. We will take advantage of that shortly.

\subsubsection{Gaussian integral representation}

We next employ the standard trick of expressing Green's functions as Gaussian integrals over commuting and anti-commuting variables. This works here without further ado, since all operators $\alpha U_d\,$, $\beta U_f\,$, $\bar\alpha U_d^{-1}$, and $\bar\beta U_f^{-1}$, are contractions (due to $|\alpha| < 1$ and $|\beta| < 1$), so that their Cayley transforms all have positive real part; cf.\ Eq.\ (\ref{eq:C-pos}).

For the Gaussian integral representation in the retarded sector, we introduce integration variables $\varphi^{\nu i} \equiv \varphi^\nu (i)$ and $\bar\varphi_{i \nu} \equiv \bar\varphi_\nu (i)$ (commuting for $\nu = 0$ and anti-commuting for $\nu = 1$) to write
\begin{equation}\label{eq:int-ret}
    \langle j \vert \left( {\textstyle{\frac{1}{2}}} (A_{\alpha U_d} + A_{\beta U_f}) \right)^{-1} \vert k \rangle = \int_{\varphi,\bar\varphi} \mathrm{e}^{-\frac{1}{2} \bar{\varphi}_\nu (A_{\alpha U_d} + A_{\beta U_f}) \varphi^\nu} \varphi^0 (j) \bar\varphi_0 (k) .
\end{equation}
To simplify the notation, we omitted the (obvious) index sums for the Hermitian form in the exponent. We note that the Gaussian integral features absolute convergence because both $A_{\alpha U_d}$ and $A_{\beta U_f}$ have positive real part.

In the advanced sector we proceed in the same way, introducing integration variables $\psi_{\ i}^\nu \equiv \psi^\nu (i)$ and $\bar\psi_{\ \nu}^i \equiv \bar\psi_\nu(i)$ to write another absolutely convergent Gaussian integral:
\begin{equation}\label{eq:int-adv}
    \langle k \vert \left( {\textstyle{\frac{1}{2}}} (A_{\bar\alpha U_d^{-1}} + A_{\bar\beta U_f^{-1}}) \right)^{-1} \vert j \rangle = \int_{\psi,\bar\psi} \mathrm{e}^{-\frac{1}{2} (-1)^{|\nu|} \psi^\nu (A_{\bar\alpha U_d^{-1}} + A_{\bar\beta U_f^{-1}}) \bar\psi_\nu} \bar\psi_0 (k) \psi^0 (j) .
\end{equation}
By compounding Eq.\ (\ref{eq:GF-Cayley}) with Eqs.\ (\ref{eq:int-ret}, \ref{eq:int-adv}) we get
\begin{eqnarray}\label{eq:GF-GI}
    &&\left\vert \langle j \vert (1- \zeta U_d U_f)^{-1} \vert k \rangle \right\vert^2 = \frac{ |\beta|^2}{|1 - \beta \mathrm{e}^{\mathrm{i} \theta_j}|^2 \; |1 - \beta \mathrm{e}^{ \mathrm{i} \theta_k}|^2} \times \\ &&\int \mathrm{e}^{-\frac{1}{2} \bar{\varphi}_\nu (A_{\alpha U_d} + A_{\beta U_f}) \varphi^\nu -\frac{1}{2} (-1)^{|\nu|} \psi^\nu (A_{\bar\alpha U_d^{-1}} + A_{\bar\beta U_f^{-1}}) \bar\psi_\nu} \varphi^0 (j) \bar\varphi_0 (k) \bar\psi_0 (k) \psi^0 (j) . \nonumber
\end{eqnarray}
The last expression explains retroactively why we have adjusted the positions of the indices $j$ and $k:$ we needed a work-around for our strict adherence to the summation convention.

\subsubsection{Taking the random-phase average}

We are now ready to carry out the disorder average. By our choice of model, the angular variables $\theta_i$ parameterizing the diagonal entries $\mathrm{e}^{\mathrm{i} \theta_i}$ of $U_f$ are independent and uniformly distributed random variables. We speak of random phases and denote the random-phase average as $\mathbb{E}(...)$. Our interest here is in the random-phase average of the squared Green's function:
\begin{equation}
    \mathbb{E} \left\vert \langle j \vert (1- \zeta U_d U_f)^{-1} \vert k \rangle \right\vert^2 .
\end{equation}
Now since our integral representations (\ref{eq:int-ret}, \ref{eq:int-adv}) are absolutely convergent, we may change the order of operations and take the random-phase average inside these integrals.

By the diagonality of $U_f$ in our basis $\vert i \rangle$ for $\mathcal{H}$, the random-phase integral factors as a product of one-dimensional integrals in that basis. The form of the integral is the same for all $i$ but for the initial and final states of the Green's function at hand. Let us first consider the case of $i = j$. By inspection of Eq.\ (\ref{eq:GF-GI}), we see that we are facing the following integral:
\begin{equation}\label{eq:phase-av}
    \frac{1}{2\pi} \int_0^{2\pi} \frac{d\theta_j} {|1 - \beta \mathrm{e}^{\mathrm{i} \theta_j}|^2}
    \exp \left( - {\textstyle{\frac{1}{2}}} \frac{1+\beta \mathrm{e}^{\mathrm{i} \theta_j}} {1-\beta \mathrm{e}^{\mathrm{i} \theta_j}} \bar\varphi_\nu(j) \varphi^\nu (j) - {\textstyle{\frac{1}{2}}} \frac{1+\bar\beta \mathrm{e}^{-\mathrm{i} \theta_j}} {1 - \bar\beta \mathrm{e}^{- \mathrm{i} \theta_j}} \bar\psi_\nu (j) \psi^\nu (j) \right) .
\end{equation}
We have been careful to insert the scalar factors $\alpha$, $\beta$ so as to arrange for $\parallel\! \alpha U_d \!\parallel < 1$, $\parallel\! \beta U_d \!\parallel < 1$, etc.; however, seeing that the absolute convergence of the integral (\ref{eq:GF-GI}) is already guaranteed by the presence of $A_{\alpha U_d}$ and $A_{\bar\alpha U_d^{-1}}$, we may now consider taking $\beta$ to the unitary limit, say $\beta \to 1$. To compute the integral (\ref{eq:phase-av}) in that limit, we observe that
\begin{equation}
    \frac{1 + \mathrm{e}^{\mathrm{i} \theta_j}} {1 - \mathrm{e}^{\mathrm{i} \theta_j}} =
    - \overline{\frac{1 + \mathrm{e}^{\mathrm{i} \theta_j}} {1 - \mathrm{e}^{\mathrm{i} \theta_j}}} = - \frac{1 + \mathrm{e}^{-\mathrm{i} \theta_j}} {1 - \mathrm{e}^{- \mathrm{i} \theta_j}}
\end{equation}
is an imaginary number, and we make the variable substitution
\begin{equation}
    x \equiv \frac{\mathrm{i}}{2} \; \frac{1 + \mathrm{e}^{\mathrm{i} \theta_j}} {1 - \mathrm{e}^{\mathrm{i} \theta_j}}\,, \quad dx = \frac{d\theta_j} {|1 -\mathrm{e}^{\mathrm{i} \theta_j}|^2} \,.
\end{equation}
Our integral (\ref{eq:phase-av}) is then seen to tend (in the limit $\beta \to 1$) to a Dirac $\delta$-distribution supported at zero:
\begin{equation}
    \frac{1}{2\pi} \int_\mathbb{R} dx\, \mathrm{e}^{-\mathrm{i} x q(j)} = \delta\big( q(j) \big) , \quad q(j) = \bar\varphi_\nu(j) \varphi^\nu (j) - \bar\psi_\nu (j) \psi^\nu (j) ;
\end{equation}
if one stops the process of taking $|\beta| \to 1$ before the limit is reached, one ends up with a regularized form of the $\delta$-distribution. Here ends the calculation for the case of $i = j$.

The situation for $i = k$ is the same (in the limit $\beta \to 1$), giving the same result. In the case of generic $i$, where $j \not= i \not= k$ and the factor $| 1 - \mathrm{e}^{\mathrm{i} \theta_i}|^{-2} = x^2 + 1/4$ is absent, we encounter a different integral:
\begin{equation}
    \frac{1}{2\pi} \int_\mathbb{R} \frac{dx}{x^2 + 1/4}\, \mathrm{e}^{-\mathrm{i} x q(i)} = \mathrm{e}^{-|q(i)|/2}, \quad q(i) = \bar\varphi_\nu(i) \varphi^\nu (i) - \bar\psi_\nu (i) \psi^\nu (i) .
\end{equation}
Here we notice a non-analyticity in $q(i) = 0$; this is rounded off when $|\beta| < 1$.

In summary, we have derived the following exact and (with proper regularization understood) rigorous formula for the random-phase average of the squared Green's function:
\begin{eqnarray}\label{eq:GF-fin}
    &&\mathbb{E} \left\vert \langle j \vert (1- \zeta U_d U_f)^{-1} \vert k \rangle \right\vert^2 = \int \mathrm{e}^{-\frac{1}{2} \bar{\varphi}_\nu (A_{\zeta U_d}) \varphi^\nu -\frac{1}{2} (-1)^{|\nu|} \psi^\nu (A_{\bar\zeta U_d^{-1}}) \bar\psi_\nu} \cr &&\hspace{1cm} \times \, \mathrm{e}^{- \frac{1}{2} \sum_{i: \, j\not= i \not= k} | q(i) |} \times \psi^0 (j) \varphi^0 (j)\, \delta \left( q(j) \right) \times \bar\varphi_0 (k) \bar\psi_0 (k)\, \delta \left( q(k) \right) .
\end{eqnarray}
Needless to say, similar formulas can be derived for other observables of the same type.

\subsubsection{Discussion of the new formula}

Our first (and rather basic) remark on Eq.\ (\ref{eq:GF-fin}) is that its integrand exhibits, in the limit of a vanishing regularization $\zeta \to 1$, the hyperbolic symmetry first pointed out by Wegner \cite{Wegner79}. Indeed, the relation $A_{U_d^{-1}} = - A_{U_d}$ makes it possible to combine the summands in the exponent into a single term:
\begin{equation}
    - \frac{1}{2} \left( A_{U_d} \right)_{\ l}^i \left( \bar\varphi_{i \nu} \varphi^{\nu l} - \bar\psi_{\ \nu} ^l \psi_{\ i}^\nu \right) \qquad (\zeta \to 1) ;
\end{equation}
we thus see that the complete integrand, with the exception of the two insertions $\psi^0 (j) \varphi^0 (j)$ and $\bar\varphi_0 (k) \bar\psi_0 (k)$, depends only on hyperbolic invariants:
\begin{equation}\label{eq:hyp-form}
    \bar\varphi_\nu(\bullet) \varphi^\nu (\bullet^\prime) - \bar\psi_\nu (\bullet^\prime) \psi^\nu (\bullet) = \bar\varphi_\nu(\bullet) \varphi^\nu (\bullet^\prime) - (-1)^{|\nu|} \psi^\nu (\bullet) \bar\psi_\nu (\bullet^\prime) .
\end{equation}
Understanding the role of the anti-commuting variables ($|\nu| = 1$) correctly, we rediscover that the global symmetry group of our problem is $G = \mathrm{U} (1,1|2) \supset \mathrm{U}(1,1) \times \mathrm{U}(2)$.

A second remark is that the present approach leading to Eq.\ (\ref{eq:GF-fin}) is dual, in a sense, to what one does in the standard Wegner-Efetov approach. Indeed, in the latter one works with the quadratic form given by $1-U_d$ (actually, $1 - U_d U_f$), whereas our Cayley transform $A_{U_d}$ puts $1-U_d$ in the denominator. Now if $U_d$ is translation-invariant, this inversion of the form amounts to a kind of $S$-duality: when a given momentum component of the field fluctuates strongly in the theory with quadratic form $1-U_d\,$, it fluctuates weakly in the dual theory with quadratic form $(1-U_d)^{-1}$, and vice versa.

Third, a salient feature of the expression (\ref{eq:GF-fin}) is the presence of two $\delta$-distributions, pinning the integration variables for $i = j$ and $i = k$ to the ``light cone'' of the hyperbolic form (\ref{eq:hyp-form}):
\begin{equation}\label{eq:LC}
    q(i) = \bar\varphi_\nu(i) \varphi^\nu (i) - \bar\psi_\nu (i) \psi^\nu (i) = 0 \quad (i=j,k) .
\end{equation}
More importantly, the factors $\prod_i \mathrm{e}^{- |q(i)|/2}$ away from $i = j, k$ put the maximum statistical weight on that very light cone. The weighting by those factors is promoted to a strong light-cone constraint (or pinning) if the deterministic factor $U_d$ induces enough collectivity in the fields $\varphi$, $\psi$ in order for $i \mapsto q(i)$ to be almost constant over a large range of $i$. (By the aforementioned $S$-duality principle, we do not expect that to happen for models in the metallic regime.) We mention in passing that the same kind of pinning comes about in the Wegner-Efetov treatment of Hamiltonian models $H = H_d + H_f$ with a random potential $H_f$ in the limit of infinite disorder strength.

Fourth, it still remains to be seen whether the present approach can lead to significant progress with a broad class of models. The success of the standard Wegner-Efetov approach derives largely from the step of introduction (by Hubbard-Stratonovich transformation) of a collective matrix field which captures the low-energy physics of diffusion (and quantum corrections to diffusion) in the metallic regime, and it is not obvious how to implement an analogous step here. One might be tempted to try bosonization to a supermatrix field constructed from bilinears in $\varphi$, $\psi$ and their conjugates. That, however, is not immediately possible, as Eq.\ (\ref{eq:GF-fin}) is not expressed in terms of local invariants and, moreover, the rank condition for the validity of superbosonization in standard form \cite{LSZ} is not satisfied.

Our fifth remark is of a philosophical nature. We recall from Sections \ref{sect:1stfail}--\ref{sect:proof} that the color-flavor transformation for models with random-phase disorder ($N = 1$) is beleaguered by the complication of unwieldy boundary correction terms. Now, what should one think of those? An optimistic aficionado of the color-flavor transformation \cite{HaakeEtAl} might have hoped that this complication is no more than an annoying technicality that can be safely ignored. On the contrary, a skeptic might suspect that the complication could be a warning signal for the possibility of another physical scenario. This author tends to be of the second opinion, based on his experience with the Chalker-Coddington network model \cite{Chalker} for the transition between plateaus of the integer quantum Hall effect in two dimensions.

For that model, the deterministic factor $U_d$ in $U = U_d U_f$ features four species of Dirac modes at low momentum \cite{CFTIQHT}. In such a situation one has the option of passing to a composite-field formulation by the procedure of non-Abelian bosonization \cite{Witten}. Now the collectivity of the composite field causes hard pinning to the light cone (\ref{eq:LC}), which has a drastic consequence: since the invariant light-cone geometry is highly anisotropic (due to the geodesic distance along null directions being zero), the field theory undergoes a novel mechanism of spontaneous symmetry breaking and rank reduction. Ultimately, this scenario leads to a description of the critical point as a conformal field theory of Wess-Zumino-Witten type, as proposed in \cite{CFTIQHT}. Yet, if one treats the Chalker-Coddington model by the type-$A$ color-flavor transformation for $N=1$ while omitting the boundary correction terms of Sections \ref{sect:fix} and \ref{sect:proof}, then one is misled to think \cite{MZ-network} that the critical point can be described by Pruisken's nonlinear $\sigma$ model.

\section{Condition of validity}\label{sect:bounds}
\setcounter{equation}{0}

In the previous section, we went on an excursion to exhibit small-$N$ issues (at the example of the type-$A$ color-flavor transformation) and their possible resolution. To complete our tour, we resume the development of Section \ref{sect:origin} (carried out for type $BD$) to state a sufficient condition for the color-flavor transformation to be valid in the form given in Eq.\ (\ref{eq:type-D}).

To begin, recall Fact 1 from the end of Section \ref{sect:killing}. In the light of it, our remaining task is to formulate and prove a criterion by which to ensure the existence of the integral on the right-hand side of Eq.\ (\ref{eq:type-D}). Let us first reflect on why such a proof is not superfluous but in fact necessary. In Section \ref{sect:proof} we saw that what does exist are the integrals for specific integrands, for example the product of the two entries in the fourth line of Table \ref{table:sections}; the left entry is a special polynomial in the $Z^{\mu\nu}$ -- in this instance a special linear combination of $(Z^{00})^l Z^{11}$ and $(Z^{00})^{l-1} Z^{10} Z^{01}$ --, which must \underline{not} be taken apart! (That example is for $N=1$, but the situation remains the same for $N > 1$.) Mathematically speaking, the integrand must be expressible by holomorphic sections of the given line bundle over the supermanifold of integration; if one replaces the integrand by just any function, the integral ceases to exist. This implies that there is still some proof work to be done.

Given that it is not obvious what conditions to impose in order for the integral on the right-hand side of Eq.\ (\ref{eq:type-D}) to exist, we will proceed in two steps. First, we will present that right-hand side in a different form, where the existence of the integral is easier to establish -- by reduction to the classical case of holomorphic discrete series representations. In the second step, we will convert the expression to the desired form.

\subsection{Even-odd factorization}

The first goal here is to re-organize the right-hand side of Eq.\ (\ref{eq:type-D}) in a way less suitable for physics applications but more suitable to discern the mathematical issue of convergence. To do so, we need to develop an adapted picture of our supermanifold of integration.

\subsubsection{Exterior-bundle picture of integration manifold}

Recall the Lie superalgebra $\mathfrak{g} = \widetilde{\mathfrak{osp}}(2n_0,2n_1)$, which was the backbone of the development in Section \ref{sect:Fock-rep}. There are two $\mathbb{Z}_2$-gradings on $\mathfrak{g} :$ in addition to the even-odd grading,
\begin{equation}
    \mathfrak{g} = \mathfrak{g}_0 \oplus \mathfrak{g}_1 \,,
\end{equation}
one has a grading
\begin{equation}
    \mathfrak{g} = \mathfrak{k} \oplus \mathfrak{p}
\end{equation}
by the eigenspaces with eigenvalue $+1$ and $-1$ (for $\mathfrak{k}$ resp.\ $\mathfrak{p}$) of the Cartan involution $X \mapsto \Sigma_3 X \Sigma_3\,$, where $\Sigma_3$ was defined in Eq.\ (\ref{eq:Sigma3}). Thus, $\mathfrak{g}$ decomposes as a direct sum of four subspaces:
\begin{equation}
    \mathfrak{g} = \mathfrak{k}_0 \oplus \mathfrak{p}_0 \oplus \mathfrak{k}_1 \oplus \mathfrak{p}_1 \,,
\end{equation}
where $\mathfrak{k}_\nu = \mathfrak{k} \cap \mathfrak{g}_\nu$ and $\mathfrak{p}_\nu = \mathfrak{p} \cap \mathfrak{g}_\nu$ ($\nu = 0, 1$). The two subspaces $\mathfrak{k}_0\,$, $\mathfrak{g}_0$ are complex Lie algebras:
\begin{equation}
    \mathfrak{gl}(n_0) \oplus \mathfrak{gl}(n_1) = \mathfrak{k}_0 \subset
    \mathfrak{g}_0 = \mathfrak{sp}(2n_0) \oplus \mathfrak{o}(2n_1) ,
\end{equation}
which exponentiate to analytic groups
\begin{equation}
    G_\mathbb{C} = \mathrm{Sp}(2n_0,\mathbb{C}) \times \mathrm{O}(2n_1,\mathbb{C}) , \quad
    K_\mathbb{C} = \mathrm{GL}(n_0,\mathbb{C}) \times \mathrm{GL}(n_1,\mathbb{C}) ,
\end{equation}
with real subgroups
\begin{equation}
    G = \mathrm{Sp}(2n_0,\mathbb{R}) \times \mathrm{O}(2n_1) , \quad
    K = \mathrm{U}(n_0) \times \mathrm{U}(n_1) .
\end{equation}
The doubly even Lie algebra $\mathfrak{k}_0 \subset \mathfrak{g}$ acts on the doubly odd space $\mathfrak{p}_1 \subset \mathfrak{g}$ by the commutator, and $K$ (as well as $K_\mathbb{C}$) acts on it by conjugation:
\begin{equation}
    K \ni k \mapsto \mathrm{Ad}(k) : \; \mathfrak{p}_1 \to \mathfrak{p}_1 \,, \; Y \mapsto k Y k^{-1} .
\end{equation}
Given that structured setting, one associates to the principal bundle $G \to G/K$ a vector bundle with standard fibre $\mathfrak{p}_1:$
\begin{equation}
    E \equiv G \times_K \mathfrak{p}_1 \,,
\end{equation}
whose points $[ g\, ; Y ] \in E$ are equivalence classes
\begin{equation}\label{eq:GEC}
    [ g\, ; Y ] \equiv [ g k^{-1} ; \mathrm{Ad}(k) Y ] \qquad (k \in K) .
\end{equation}
Now the supermanifold of integration in Eq.\ (\ref{eq:type-D}) is modeled on the exterior bundle $\bigwedge E$ by a universal construction: choosing any basis $\{ e_\alpha \}$ of $\mathfrak{p}_1$ with dual basis $\{ f^\alpha \}$, and expressing the generic fiber element $Y$ as $Y = e_\alpha \otimes f^\alpha$, one re-interprets the matrix elements $f^\alpha$ as generators $f^\alpha \equiv \xi^\alpha$ of a Grassmann algebra $\wedge(\mathfrak{p}_1^\ast)$. The supermanifold so defined carries various group actions derived from the underlying Lie superalgebra $\mathfrak{g}$ -- by that token, it is the same as a Riemannian symmetric superspace of type $C{\rm I} \vert D{\rm I\!I\!I}$ \cite{suprev}.

Let us rephrase this in physics language. The standard ``rational'' parametrization of the Wegner-Efetov supermatrix $Q = T \Sigma_3 T^{-1}$ is by $T$ as in Eq.\ (\ref{eq:rat-T}):
\begin{equation}
    Q = T \Sigma_3 T^{-1} =
    \begin{pmatrix}
    (1 + Z \widetilde{Z})(1 - Z \widetilde{Z})^{-1}
    &2 Z (1 - \widetilde{Z} Z)^{-1} \cr
    -2 \widetilde{Z} (1 - Z \widetilde{Z})^{-1}
    &- (1 + \widetilde{Z} Z)(1 - \widetilde{Z} Z)^{-1}
    \end{pmatrix} .
\end{equation}
Our intention here is to re-parameterize it by two factors, namely $g \in G$ and the doubly odd supermatrix $Y$:
\begin{equation}
    Q = g\, \mathrm{e}^Y \Sigma_3\, \mathrm{e}^{-Y} g^{-1} .
\end{equation}
That parametrization comes with a redundancy which is quantified by gauge transformations
\begin{equation}
    g \mapsto g k^{-1} , \quad Y \mapsto k Y k^{-1} \quad (k \in K) ,
\end{equation}
whose gauge equivalence classes are the points of $G \times_K \mathfrak{p}_1\,$; cf.\ Eq.\ (\ref{eq:GEC}).

\subsubsection{Factorization of integration measure}

Much of the difficulty of seeing through the intricacies of Eq.\ (\ref{eq:type-D}) is caused by the mixing of even with odd variables in the Berezin integration form $D\mu(Z, \widetilde{Z})$ of Eq.\ (\ref{eq:invt-Dm}). We shall now see that the even-odd factorization $T = g\, \mathrm{e}^Y$ separates the variables.

To re-express the invariant Berezin integration $D\mu$ according to $T = g\, \mathrm{e}^{Y}$, we need the super-Jacobian of the transformation. This can be read off from the Cartan-Maurer form $T^{-1} dT$ projected to $\mathfrak{p}:$
\begin{equation}
    (T^{-1} dT)_\mathfrak{p} = \left( \mathrm{e}^{-Y} g^{-1} d (g \, \mathrm{e}^Y) \right)_\mathfrak{p} \,.
\end{equation}
By the doubly graded structure of $\mathfrak{g}\,$, the projected Cartan-Maurer form breaks up into three terms:
\begin{equation}
    (T^{-1} dT)_\mathfrak{p} = \left( \mathrm{e}^{-Y} d\, \mathrm{e}^Y \right)_{\mathfrak{p}_1}
    + \cosh \mathrm{ad}(Y)\, (g^{-1} d g)_{\mathfrak{p}_0} - \sinh \mathrm{ad}(Y)\, (g^{-1} d g)_{\mathfrak{k}_0} \,,
\end{equation}
where $\mathrm{ad}(Y) \bullet \equiv [Y,\bullet]$. The third summand can be gauged away and, in any event, does not contribute to our Jacobian, as it is off-diagonal (even-to-odd) with no counterpart (odd-to-even). The inner term $(g^{-1} d g)_{\mathfrak{p}_0}$ of the second summand contributes the Jacobian for the $G$-invariant measure $dg_K$ on $G/K$. The first summand is
\begin{equation}
    \left( \mathrm{e}^{-Y} d\, \mathrm{e}^Y \right)_{\mathfrak{p}_1} = \left( \frac{1 - \mathrm{e}^{-\mathrm{ad}(Y)}}{\mathrm{ad}(Y)} \, dY \right)_{\mathfrak{p}_1} = \frac{\sinh \ \mathrm{ad}(Y)} {\mathrm{ad}(Y)} \, dY .
\end{equation}
Thus the $\mathfrak{g}$-invariant Berezin integration form (\ref{eq:invt-Dm}) is re-computed as
\begin{equation}
    D\mu = dg_K \, \partial_Y \circ \Omega(Y) ,
\end{equation}
where $\partial_Y = \prod_\alpha \partial / \partial\xi^\alpha$ is the flat Berezin form for the odd vector space $\mathfrak{p}_1\,$, and $\Omega$ is a function of only the Grassmann variables $\xi^\alpha:$
\begin{equation}
    \Omega(Y = e_\alpha \xi^\alpha) = \mathrm{Det}^{-1} \frac{\sinh \, \mathrm{ad}(Y)}{\mathrm{ad}(Y)} \big\vert_{\mathfrak{p}_1 \to \mathfrak{p}_1} \; \mathrm{Det} \cosh \mathrm{ad}(Y) \big\vert_{\mathfrak{p}_0 \to \mathfrak{p}_0} \,.
\end{equation}
Note that $\Omega$ is a superdeterminant: the factor from the even sector ($\mathfrak{p}_0 \to \mathfrak{p}_0$) appears in the numerator, that from the odd sector ($\mathfrak{p}_1 \to \mathfrak{p}_1$) goes in the denominator. It is possible to evaluate the expression for $\Omega$ further, but this will not be needed for now. We note that $\Omega$ is $K$-invariant:
\begin{equation}\label{eq:Om-Kinvt}
    \Omega\big( \mathrm{Ad}(k) Y \big) = \Omega(Y) .
\end{equation}

\subsubsection{Lift to the Fock representation}

Recall from Section \ref{sect:Fock-rep} the Fock-space representation $X \mapsto \widehat{X}$ of $\mathfrak{g} = \widetilde{\mathfrak{osp}}(2n_0 , 2n_1)$. We shall now use that representation to transfer the factorized supermatrix $Q = T \Sigma_3 T^{-1} = g \, \mathrm{e}^Y \Sigma_3 \, \mathrm{e}^{-Y} g^{-1}$ to an operator on the Fock space with vacuum $\vert 0 \rangle$:
\begin{equation}\label{eq:Q-Fock}
    Q \to \mathcal{D}(g)\, \mathcal{D}(\mathrm{e}^Y)\, \vert 0 \rangle \langle 0 \vert \,\mathcal{D}(\mathrm{e}^{-Y})\, \mathcal{D}(g^{-1}) ,
\end{equation}
This needs a few comments. First of all, the representation of the odd factor is immediate by $\mathcal{D}( \mathrm{e}^Y ) \equiv \mathrm{e}^{\widehat{Y}}$, since the Taylor series for the exponential of $\widehat{Y}$ is finite. Second, the representation $\mathcal{D}(g)$ of the even factor $g \in G$ has two parts associated with the factors of the direct product $G = \mathrm{Sp}(2n_0, \mathbb{R}) \times \mathrm{SO}(2n_1)$. The part from the odd-odd (or fermion-fermion) sector $\mathrm{SO}(2n_1)$ can still be defined by exponentiation, say $\mathcal{D} (\mathrm{e}^{X_{\rm FF}}) \equiv \mathrm{e}^{\widehat{X}_{\rm FF}}$; there is just a slight complication due to the homotopy group $\pi_1 \big( \mathrm{SO}(2n_1) \big)$ being nontrivial: for odd values of the number $N$ of colors, the exponentiated representation turns out to be a representation of the simply connected group $\mathrm{Spin} (2n_1)$ covering $\mathrm{SO} m(2n_1)$.

More analytical effort is needed to deal with the boson-boson sector, $\mathrm{Sp}(2n_0,\mathbb{R})$, as the exponentiated operator $\mathcal{D}(\mathrm{e}^{X_{\rm BB}}) \stackrel{?}{=} \mathrm{e}^{\widehat{X}_{\rm BB}}$ does not exist for $X_{\rm BB} \in \mathfrak{sp}(2n_0)$ in general. As we remarked earlier (at the beginning of Section \ref{sect:strategy}), what does exponentiate is the representation on a half-space, leading to the representation of a contraction semigroup in $\mathrm{Sp}(2n_0,\mathbb{C})$. Now the real form $\mathfrak{sp}(2n_0,\mathbb{R})$ of the complex Lie algebra $\mathfrak{sp}(2n_0)$ is contained in the boundary of that half-space, making it possible to construct the exponentiated representation by a limit procedure. The final outcome of such analysis is mathematically well-established as the Segal-Shale-Weil representation (or a generalization thereof) of the metaplectic group $\mathrm{Mp}(2n_0) \stackrel{2:1}{\longrightarrow} \mathrm{Sp}(2n_0, \mathbb{R})$, and we shall now take it for granted and use it. (The author's favorite reference on the subject is \cite{Howe-oscrep}.)

Our main idea is that by integrating the operator (\ref{eq:Q-Fock}) one obtains another (tentative) expression for the coherent-state transition amplitudes of the color-singlet projection operator $P$ in Eqs.\ (\ref{eq:intro-P}, \ref{eq:3.16}):
\begin{equation}\label{eq:P-New}
    \langle \psi \vert P \vert \psi^\prime \rangle \stackrel{?}{=} \tilde{c}_{N}^{(BD)}  \int D\mu \ \langle \psi \vert \mathcal{D}(g) \mathcal{D}(\mathrm{e}^Y) \vert 0 \rangle \langle 0 \vert \mathcal{D}(\mathrm{e}^{-Y}) \mathcal{D}(g^{-1}) \vert \psi^\prime \rangle .
\end{equation}
The advantage over the old expression (\ref{eq:3.16}) is that for the new expression (\ref{eq:P-New}) it will be easier to decide whether or not the integral with invariant measure $D\mu$ converges.

A major simplification occurs because the even-odd factorization lets us carry out the Grassmann-Fermi integral first. Let
\begin{equation}
    \partial_Y \circ \Omega(Y) \ \mathcal{D}(\mathrm{e}^Y) \vert 0 \rangle \langle 0 \vert \mathcal{D}(\mathrm{e}^{-Y}) \equiv \Pi .
\end{equation}
The resulting operator $\Pi$ is the ``density matrix'' for a mixed state made from color-neutral boson-fermion pairs (created and annihilated by $\widehat{Y}$).

Now recall Eq.\ (\ref{eq:Om-Kinvt}) and note the consequence that the Berezin form $\partial_Y \circ \Omega(Y)$ is $K$-invariant. Since the Fock vacuum carries a one-dimensional $K$-representation, it follows that the vacuum projector $\vert 0 \rangle \langle 0 \vert$ is $K$-invariant and so is $\Pi$:
\begin{equation}
    \mathcal{D}(k)\, \Pi\, \mathcal{D}(k^{-1}) = \Pi \quad (k \in K) .
\end{equation}
Therefore, the integrand of the remaining integral,
\begin{equation}
    \langle \psi \vert P \vert \psi^\prime \rangle \stackrel{?}{=} \tilde{c}_{N}^{(BD)}  \int_{G/K} \!\!\!\! dg_K \, \langle \psi \vert \mathcal{D}(g)\, \Pi\, \mathcal{D}(g^{-1}) \vert \psi^\prime \rangle ,
\end{equation}
is well-defined as a function on the quotient $G/K$.

The $K$-invariant mixed-state density operator $\Pi$ can be expanded as a sum over irreducible $K$-representations indexed by, say, a highest weight $\lambda_K$ for the representation and another label, $m$, running through a set of basis vectors for that representation. Using Dirac's notation, we write
\begin{equation}
    \Pi = \sum \vert \lambda_K , m \rangle \langle \lambda_K , m \vert \,.
\end{equation}
The first term in the sum is the projector $\vert 0 \rangle \langle 0 \vert$ on the Fock vacuum. The other states in the sum result from the vacuum by adding color-neutral pairs, each made by pairing one boson with one fermion. It should be noted that the latter $K$-representations are, in general, not highest-weight for the action of $G$.

Now recall that the Lie group $G$ factors into even-even and odd-odd parts:
\begin{equation}
    G = G_0 \times G_1 \,, \quad G_0 \equiv \mathrm{Sp}(2n_0,\mathbb{R}) , \quad G_1 \equiv \mathrm{SO}(2n_1) ,
\end{equation}
and
\begin{equation}
    K = K_0 \times K_1 \,, \quad K_0 \equiv \mathrm{U}(n_0) , \quad K_1 \equiv \mathrm{U}(n_1) .
\end{equation}
Thus we have $G/K = M_0 \times M_1$ with $M_0 = G_0 / K_0$ and $M_1 = G_1 / K_1\,$. Correspondingly, the $G$-invariant integration measure for $G/K$ factors:
\begin{equation}
    dg_K = dg_{K,0} \times dg_{K,1} \,.
\end{equation}
Now the integral over the compact symmetric space $M_1$ with invariant measure $dg_{K,1}$ always exists for a bounded integrand such as that in Eq.\ (\ref{eq:P-New}). What remains in question is the integral over the noncompact symmetric space $M_0$ with invariant measure $dg_{K,0}\,$.

The latter integral can be pulled back to an integral over $\mathrm{Mp}(2n_0) \equiv \widetilde{G}_0$ with Haar measure $d g_0$ by the projection $\pi : \; \widetilde{G}_0 \to \widetilde{G}_0 / \widetilde{K}_0 = G_0 / K_0 = M_0$ with compact fiber. We thus see that the question of (a sufficient condition for the) existence of the integral (\ref{eq:P-New}) reduces to the existence question for integrals of products of two matrix coefficients,
\begin{equation}\label{eq:L2-Mp}
    \int_{\widetilde{G}_0} \!\! dg_0 \, \langle \bullet \vert \mathcal{D}(g_0) \vert \bullet^\prime \rangle \langle \bullet^{\prime\prime} \vert \mathcal{D}(g_0^{-1}) \vert \bullet^{\prime\prime\prime} \rangle ,
\end{equation}
taken between any state vectors of the (reducible) $\mathrm{Mp}(2n_0)$-representation on Fock space.

The decomposition of the Fock-space representation into irreducibles for $\mathrm{Mp}(2n_0)$ is well understood \cite{Howe-Schur}: all of these irreps are of a type known as \emph{holomorphic discrete series}; c.f.\ \cite{stable}. For that type, the answer to the question of existence of the integrals (\ref{eq:L2-Mp}) is as follows.

\subsubsection{$L^2$-integrability of holomorphic discrete series}

We shall now state and apply the relevant integrability criterion as given in \cite{stable}. To do so, we need to explain some notation first. This will look a little strange, as our Fock vacuum is most naturally a \emph{lowest}-weight vector, whereas standard representation theory employs the language of \emph{highest} weights; therefore, to get a match, we must arrange for pair creation to correspond to negative roots and pair annihilation to positive roots.

Taking the Cartan subalgebra $\mathfrak{t} \subset \mathfrak{g}_0$ to be spanned by diagonal matrices
\begin{equation}
    h = \mathrm{diag} (h^1, \ldots, h^{n_0} \,; - h^1, \ldots, - h^{n_0}) ,
\end{equation}
we define a basis $f^\mu$ of linear functions on $\mathfrak{t}$ by
\begin{equation}
    f^\mu(h) = h^\mu \quad (\mu = 1, \ldots, n_0) .
\end{equation}
The positive noncompact roots ($\Delta_{\mathfrak{p}}^+$) and positive compact roots ($\Delta_{\mathfrak{k}}^+$) for the adjoint action of $\mathfrak{t}$ on $\mathfrak{g}_0$ then are:
\begin{equation}
    \Delta_{\mathfrak{p}}^+ : \; f^\mu + f^\nu \quad (\mu \leq \nu) , \quad
    \Delta_{\mathfrak{k}}^+ : \; f^\mu - f^\nu \quad (\mu < \nu) .
\end{equation}
Taking the union of sets, $\Delta^+ \equiv \Delta_{\mathfrak{p}}^+ \cup \Delta_{\mathfrak{k}}^+$, the half-sum of positive roots is
\begin{equation}
    \delta = \frac{1}{2} \sum_{\alpha \in \Delta^+} \alpha = n_0 f^1 + (n_0-1) f^2 + \ldots + 2 f^{n_0-1} + f^{n_0} .
\end{equation}
With these conventions, one verifies from the Fock-space representation given in Eq.\ (\ref{eq:g-rep}) that our Fock vacuum is a highest-weight vector for $\mathfrak{g}_0$ with weight
\begin{equation}\label{eq:vac-lam}
    \lambda_0 = - \frac{N}{2} \sum_{\mu=1}^{n_0} f^\mu .
\end{equation}
The half-integrality of $\lambda_0$ for odd $N$ signals that the vacuum representation is ``square-root'' or double-valued as a representation of $G_0 = \mathrm{Sp} (2n_0,\mathbb{R})$.

Now Theorem 6.6 of \cite{stable} states the following: if $\langle \bullet \vert \mathcal{D}(g_0) \vert \bullet^\prime \rangle$ are the matrix coefficients for a $\widetilde{G}_0$-representation of the holomorphic discrete series with highest weight $\lambda$, then they are in $L^2(\widetilde{G}_0)$ if and only if the highest weight satisfies
\begin{equation}\label{eq:thm6.6}
    \langle \lambda + \delta\, , \alpha \rangle < 0 \quad ({\rm for \; all}\; \alpha \in \Delta_{\mathfrak{p}}^+) ,
\end{equation}
where $\langle \cdot , \cdot \rangle$ denotes the inner product of $\mathfrak{t}^\ast$ induced by the Cartan-Killing form of $\mathfrak{g}_0\,$. In the case of our vacuum-based representation with highest weight $\lambda = \lambda_0$ given in (\ref{eq:vac-lam}), a simple computation shows that this condition amounts to $N \geq 2 n_0 + 1$.

The vacuum-based representation (spanned by adding color-neutral boson pairs to the vacuum) is just one of many $\widetilde{G}_0$-representations contained in our Fock space of colored bosons and fermions. The other representations that occur are still in the holomorphic discrete series and hence are covered by the theorem above. Their highest-weight vectors are constructed by adding color-neutral boson-fermion pairs to the vacuum and then removing by orthogonal projection any component in the vacuum-based representation. Since the positive noncompact roots $\alpha \in \Delta_{\mathfrak{p}}^+$ correspond to boson pair annihilation, the root-opposite process of (boson-fermion) pair creation modifies the highest weight ($\lambda_0 \to \lambda$) by an amount
\begin{equation}
    \langle \lambda - \lambda_0 , \alpha \rangle < 0 \quad (\alpha \in \Delta_{\mathfrak{p}}^+) ,
\end{equation}
which makes the scalar product $\langle \lambda + \delta \,, \alpha \rangle$ more negative. Therefore, if the vacuum highest weight $\lambda_0$ satisfies the integrability condition (\ref{eq:thm6.6}), then so do the highest weights of all the other $\widetilde{G}_0$-representations in Fock space. Thus the condition $N \geq 2 n_0 + 1$ guarantees the convergence of the integral (\ref{eq:P-New}) and ultimately of the color-flavor transformation (\ref{eq:type-D}). 


\bigskip\noindent\textbf{Acknowledgments.} The author credits Peter Sarnak (1997) for the remark that the color-flavor transformation might be seen as a corollary of Howe duality. He acknowledges Gabriele  Campagnano (2004) for raising some of the issues with $N=1$ that were discussed and brought to clarification in Section \ref{sect:TPfctn}.

\end{document}